# Cluster lens reconstruction using only observed local data – an improved finite-field inversion technique

Stella Seitz & Peter Schneider
Max-Planck-Institut für Astrophysik
Postfach 1523
D-85740 Garching, Germany

**Abstract:**

Gravitational light deflection can distort the images of distant sources by its tidal effects. The population of faint blue galaxies is at sufficiently high redshift so that their images are distorted near foreground clusters, with giant luminous arcs being the most spectacular evidence for this effect. Much weaker distortions, however, can observationally be detected by a statistical analysis of the numerous faint galaxy images, as first demonstrated by Tyson, Valdes & Wenk. This distortion effect can be used as a *quantitative tool* for the reconstruction of the surface mass density of galaxy clusters with appropriate redshifts, as was demonstrated by Kaiser & Squires. They have derived an explicit equation for this surface mass density in terms of its tidal field.

    The reconstruction formula by Kaiser & Squires must be modified because of two effects: in its original form it applies only to weak lenses, and hence must be generalized to account for stronger lensing effects. Second, due to the nature of the inversion formula, it produces boundary artefacts (or biases) if applied to real data which are confined to a finite field on the sky. We discuss several possibilities to obtain inversion formulae which are exact for ideal data on a finite data field (CCD). We demonstrate that there exists an infinite number of such finite-field inversion formulae, which differ in their sensitivity to observational effects (such as noise, intrinsic ellipticities of the sources, etc.). We show that, using two simple conditions, one can uniquely specify a finite-field formula which in a well-defined sense minimizes the sensitivity to observational effects. We then use synthetic data to compare the quality of our new reconstruction method with that of previous finite-field inversion techniques and of the nonlinear generalization of the Kaiser & Squires method. This analysis demonstrates that our new inversion method is superior to the other previously considered finite-field inversions, and only slightly more noisy than the Kaiser & Squires inversion, but in contrast to the latter, it lacks the boundary artefacts. We shall discuss that the lack of boundary artefacts and the slightly increased noise have the same origin, and that every finite-field reconstruction must be more noisy than that obtained from the Kaiser & Squires method. Furthermore, the rms deviations of the reconstructed density field from the input surface mass density are fairly homogeneously distributed over the field. We therefore conclude that the new method developed here is the best inversion formula yet found.

# 1 Introduction

Weak distortions of images of faint, high-redshift galaxies due to the gravitational light deflection caused by a foreground cluster can be used as a probe for the mass distribution of the cluster (e.g., Webster 1985). This effect manifests itself most visibly in the giant luminous arcs, where the image distortion is very strong (see the recent review by Fort & Mellier 1994). It is, of course, more likely that a galaxy image is less dramatically distorted, and the first of those more weakly distorted images (called arclets) were found by Fort et al. (1988). With one giant luminous arc in A370 and several arclets, the mass distribution of the cluster could be much better constrained than from single arcs alone (e.g., Grossman & Narayan 1989, Kneib et al 1993). Even weaker distortions, not obviously seen in individual galaxy images but clearly detectable statistically were verified by Tyson, Valdes & Wenk (1990). Such observations can probe the cluster mass distribution to much larger angular scales from the cluster center than accessible with giant luminous arcs (Bonnet, Mellier & Fort 1994).

The distortion of images can be calculated, loosely speaking, as a convolution of the surface mass density with a (known) weighting funtion. In their pioneering paper, Kaiser & Squires (1993, hereafter KS) have shown that this relation can be inverted, i.e., the surface mass density distribution can be obtained as a convolution of the distortion field with a (known) kernel (earlier qualitative work on the determination of cluster mass distributions from weak distortions include Kochanek 1990; Miralda-Escudé 1991). In other words, if the distortion field can be measured with sufficient accuracy (i.e., if the observations extend to sufficiently faint limits, so that the number density of faint galaxies is high enough), a density map of the corresponding cluster can be constructed. In fact, this method has already been applied to a number of clusters (Fahlman et al. 1994; Smail et al. 1995; Kaiser et al. 1994, henceforth KSFWB), and in particular for the cluster MS1224 has led to a mass estimate which is significantly larger than estimates from a dynamical study of this cluster (Carlberg, Yee & Ellingson 1995). These first applications have amply demonstrated the potential power of this new method for investigating clusters, which is independent of assumptions about the physical or dynamical state of the cluster galaxies or the intracluster gas.

Nevertheless, the application of the KS method is far from being trivial. The difficulties with it can roughly be split into two parts: first, and certainly most serious, is the problem of extracting the distortion field from observations. It is essential for the method that the observations are taken at very good seeing, since the typical size of the galaxies – at the faint levels for which their number density is sufficiently large – is comparable to the seeing disk, so that the distortion signal is diluted. Other effects, like telescope tracking errors, lead to a non-trivial point spread function whose anisotropy generates an artifical distortion. These problems can be overcome, as discussed in detail by Bonnet & Mellier (1995), Kaiser, Squires & Broadhurst (1995) and Mould et al. (1994), and are not the subject of this paper.

The second set of problems is more of a theoretical nature. The KS inversion method relates the 'shear' of the mass distribution – which depends linearly on the surface mass density – to the mass distribution. However, the shear is not an observable. As was pointed out in Schneider & Seitz (1995, hereafter Paper I), the observable is a combination of the shear and the surface mass density; only in the weak-lensing regime can the shear be obtained from observable quantities. This fact complicates the inversion process due to its nonlinearity. In Seitz & Schneider (1995, hereafter Paper II) it was shown that a general two-dimensional nonlinear reconstruction is possible, so that the KS method can be extended towards the inner parts of the lensing cluster. This method, though slightly more complicated than the original KS method, works as well as the KS inversion in those regimes where both can be applied. A second problem with the KS method is the fact that the inversion formula in principle requires data on the entire sky, whereas the observed field around a cluster is always bounded. Hence, the shear field has to be extrapolated outside the data field. The simplest such 'extrapolation' is setting the shear to zero outside the data field. This, however, leads to serious 'boundary effects' in the reconstructed density field. Alternatively, one can fit a model function to the shear field in the outer parts of the data field and use this model function for the extrapolation, as has been done in Paper II. Using a sample of cluster models drawn from a high-resolution CDM simulation, it was shown by Bartelmann (1995) that this extrapolation yields in fact fairly accurate estimates of the cluster masses, and the reconstructed density fields in Paper II show that the boundary artefacts obtained from the original KS inversion can be reduced significantly by this extrapolation. Nevertheless, both of these two extrapolations mean 'inventing data' and should thus be avoided.

It was shown by Schneider (1995, hereafter Paper III), using a relation between the gradient of the surface mass density and a combination of derivatives of the shear which was obtained by Kaiser (1995), that a cluster inversion formula can be obtained which explicitly makes use only of data on a finite field. This formula was obtained by a carefully chosen average of line integrals of the gradient of the surface mass density over the data field. Subsequently, KSFWB have generalized the ideas of Paper III to obtain different sets of such finite-field inversion equations, one of which was explored in detail by Bartelmann (1995). We shall summarize these finite-field methods in Sect. 3 below, where it will become clear that one has a huge freedom for constructing such inversion formulae. It thus remains questionable which of these is the 'best' one – all of them yield exact results for perfect data, but they behave differently with respect to noise, both from observations and from the fact that the faint galaxies have an intrinsic ellipticity distribution.

In this paper we shall reconsider the problem of finding a finite-field inversion equation. The notation and the nature of the problem are introduced in Sect. 2. In Sect. 3 we derive families of finite-field inversions based on 'averaging line integrals' of the gradient of the surface mass density, of which the ones of Paper III and of Bartelmann (1995) are special cases; this section summarizes the previous attempts for constructing finite-field inversion methods and generalizes them slightly. In Sect. 4 we derive a finite-field inversion formula which is *uniquely specified* by two well-motivated requirements. In Sect. 5 we compare the various inversion techniques (KS, the one of Paper III, that from Bartelmann 1995, and our new one) using simulated distortion fields. It turns out that our new inversion method is superior to the other finite-field methods, and that it has (at worst) only marginally worse noise properties than the KS method, but the advantage that it is free from (systematic) boundary artefacts. The relative noise properties of our new method and KS depend strongly on the underlying mass distribution; if it extends to near the boundary of the data field, the noise properties of our inversion are better than that of KS. Also, we show that the deviations of our reconstructed mass profile from the original one are statistically homogeneously distributed over the data field, in contrast to the other reconstruction formulae. We thus conclude that our new inversion technique is the best found up to now, since the slightly increased noise relative to KS in some situations is well compensated by the lack of systematic boundary artefacts.

Throughout this paper we consider the possibility that the lens is not weak, so the linear approximation of the shear is not used. Hence, if we apply the KS method, we always do so by the iterative procedure developed in Paper II. However, we assume that the lenses are non-critical, so that the local degeneracy found in Paper I does not occur. Furthermore, as in the previous papers, we will assume that all sources are at the same redshift; this assumption simplifies the inversion process considerably, but is not essential for the application of our inversion technique. We shall deal with a redshift distribution of the sources in a later paper. If we were to consider only weak lenses, and thus use the linear approximation, then a redshift distribution of the sources causes no additional technical problems.

## 2 Inversion methods

We use the same notation as in Papers I–III. Briefly, we define the deflection potential

$$\psi(\boldsymbol{\theta}) = \frac{1}{\pi} \int_{\mathbb{R}^2} \mathrm{d}^2\theta' \, \kappa(\boldsymbol{\theta}') \ln \left|\boldsymbol{\theta} - \boldsymbol{\theta}'\right| \tag{2.1}$$

in terms of the dimensionless surface mass density $\kappa(\boldsymbol{\theta})$. The linearized lens mapping $\boldsymbol{\beta} = \boldsymbol{\theta} - \nabla\psi(\boldsymbol{\theta})$, which describes the distortion of small sources, is $\mathrm{d}\boldsymbol{\beta} = A(\boldsymbol{\theta})\,\mathrm{d}\boldsymbol{\theta}$, where $\boldsymbol{\beta}$ denotes the angular position on the source sphere, and

$$A(\boldsymbol{\theta}) = \begin{pmatrix} 1 - \kappa + \gamma_1 & +\gamma_2 \\ +\gamma_2 & 1 - \kappa - \gamma_1 \end{pmatrix} \tag{2.2}$$

is the Jacobian matrix of the lens mapping, which is related to the deflection potential $\psi$ through

$$\begin{aligned}
\kappa(\boldsymbol{\theta}) &= \frac{1}{2}\nabla^2\psi(\boldsymbol{\theta}) \quad , \\
\gamma_1(\boldsymbol{\theta}) &= \frac{1}{2}\left(\psi_{,22} - \psi_{,11}\right) \quad , \\
\gamma_2(\boldsymbol{\theta}) &= -\psi_{,12} \quad ,
\end{aligned} \tag{2.3}$$

where indices separated by a comma denote partial derivatives with respect to $\theta_i$. For details concerning these lensing relations, see Schneider, Ehlers & Falco (1992). Combining (2.1 & 2.3), and defining the complex shear $\gamma(\boldsymbol{\theta}) = \gamma_1(\boldsymbol{\theta}) + \mathrm{i}\gamma_2(\boldsymbol{\theta})$, one obtains

$$\gamma(\boldsymbol{\theta}) = \frac{1}{\pi} \int_{\mathbb{R}^2} \mathrm{d}^2\theta' \, \mathcal{D}(\boldsymbol{\theta} - \boldsymbol{\theta}') \kappa(\boldsymbol{\theta}') \quad , \tag{2.4}$$

where

$$\mathcal{D}(\boldsymbol{\theta}) = \frac{\theta_1^2 - \theta_2^2 + 2\mathrm{i}\theta_1\theta_2}{|\boldsymbol{\theta}|^4} \equiv \mathcal{D}_1(\boldsymbol{\theta}) + \mathrm{i}\mathcal{D}_2(\boldsymbol{\theta}) \tag{2.5}$$

is a complex kernel. It was shown in KS that (2.5) can be inverted to yield

$$\kappa(\boldsymbol{\theta}) = \frac{1}{\pi} \int_{\mathbb{R}^2} \mathrm{d}^2\theta' \, \mathcal{Re}\big[\mathcal{D}^*(\boldsymbol{\theta} - \boldsymbol{\theta}') \gamma(\boldsymbol{\theta}')\big] \quad . \tag{2.6a}$$

Using partial integration and the assumption that the shear vanishes at infinity yields a different form of the inversion integral,

$$\kappa(\boldsymbol{\theta}) = \int_{\mathbb{R}^2} \mathrm{d}^2\theta' \, \mathbf{H}^{\mathrm{KS}}(\boldsymbol{\theta}', \boldsymbol{\theta}) \begin{pmatrix} -\gamma_{1,1}(\boldsymbol{\theta}') - \gamma_{2,2}(\boldsymbol{\theta}') \\ -\gamma_{2,1}(\boldsymbol{\theta}') + \gamma_{1,2}(\boldsymbol{\theta}') \end{pmatrix} \quad , \tag{2.6b}$$

with

$$\mathbf{H}^{\mathrm{KS}}(\boldsymbol{\theta}', \boldsymbol{\theta}) = \frac{1}{2\pi} \frac{\boldsymbol{\theta} - \boldsymbol{\theta}'}{|\boldsymbol{\theta} - \boldsymbol{\theta}'|^2} = \nabla_{\boldsymbol{\theta}'} \left( -\frac{1}{2\pi} \ln |\boldsymbol{\theta} - \boldsymbol{\theta}'| \right) \quad ; \tag{2.6c}$$

hence, the surface mass density $\kappa$ at a position $\boldsymbol{\theta}$ can be obtained by convolving the deflection angle of a point mass with negative mass $-1/2$ positioned at $\boldsymbol{\theta}$ with the derivative of the shear field $\gamma$.

As mentioned in the introduction, there are two fundamental problems associated with (2.6a). First, the shear $\gamma(\boldsymbol{\theta})$ cannot be obtained from image distortions, but what can be determined is the *(complex) distortion*

$$\delta = \frac{2\gamma(1-\kappa)}{(1-\kappa)^2 + |\gamma|^2} \quad ; \tag{2.7}$$

see Paper I and Miralda-Escude 1991 for details. If we assume (as we shall do in the rest of the paper) that the cluster is non-critical (i.e., $\det A(\boldsymbol{\theta}) > 0$ for all $\boldsymbol{\theta}$), then the quantity

$$g = \frac{\gamma}{(1-\kappa)} = \left( \frac{1 - \sqrt{1 - |\delta|^2}}{|\delta|^2} \right) \delta \tag{2.8}$$

is also an observable. Combining (2.8) with (2.6a), we obtain

$$\kappa(\boldsymbol{\theta}) = \frac{1}{\pi} \int_{\mathbb{R}^2} \mathrm{d}^2\theta' \, \big[1 - \kappa(\boldsymbol{\theta}')\big] \mathcal{Re}\big[\mathcal{D}^*(\boldsymbol{\theta} - \boldsymbol{\theta}') g(\boldsymbol{\theta}')\big] \quad , \tag{2.9}$$

i.e., an integral equation for $\kappa(\boldsymbol{\theta})$, which can easily be solved iteratively, as shown in Paper II. A second problem associated with (2.6) and (2.9) is the range of integration: since data are available only on a finite field, given by the size of the CCD, these equations can only be applied to data if a 'guess' is made for $g$ outside the data field $\mathcal{U}$ (where $\mathcal{U}$ is that region within which we can determine $g$ from observations) . Typically, $g$ and thus $\gamma$ is set to zero outside the data field[1], so that

$$\kappa^{\mathrm{KS},1}(\boldsymbol{\theta}) := \frac{1}{\pi} \int_{\mathcal{U}} \mathrm{d}^2\theta' \, \mathcal{R}\mathrm{e}\left[\mathcal{D}^*(\boldsymbol{\theta} - \boldsymbol{\theta}')\,\gamma(\boldsymbol{\theta}')\right] \quad . \tag{2.10a}$$

This so called 'KS'-estimate suffers from boundary artefacts, whose amplitude depends on form and size of the data field and on the mass distribution. Alternatively, one can try to extrapolate $g$ outside the field, which is reasonable since the shear cannot decrease faster than the shear of a point mass. Such an extrapolation was successfully applied to a numerical model cluster in Paper II, and in a more systematic study Bartelmann (1995) has demonstrated that the extrapolation method yields quite accurate results for the mass within circular apertures. One can obtain a different estimate for $\kappa$ from (2.6b) by setting the derivatives of $\gamma$ to zero outside the data field

$$\kappa^{\mathrm{KS},2}(\boldsymbol{\theta}) := \int_{\mathcal{U}} \mathrm{d}^2\theta' \, \mathbf{H}^{\mathrm{KS}}(\boldsymbol{\theta}', \boldsymbol{\theta}) \begin{pmatrix} -\gamma_{1,1}(\boldsymbol{\theta}') - \gamma_{2,2}(\boldsymbol{\theta}') \\ -\gamma_{2,1}(\boldsymbol{\theta}') + \gamma_{1,2}(\boldsymbol{\theta}') \end{pmatrix} \quad . \tag{2.10b}$$

This estimate is expected to differ from that in (2.10a) in the systematic boundary artefacts as well as in the noise properties.

In order to determine mass profiles, one should aim for an inversion of (2.4) which makes use only of the observed data on $\mathcal{U}$ and which yields an exact reconstruction for perfect data. As was shown by KS, the surface mass density can be determined only up to an additive constant, *if the shear $\gamma$ is assumed to be measurable.*[2] Choosing this constant to be the (unknown) mean surface mass density $\bar\kappa$ over the data field, one may want to search for a kernel $\tilde{\mathcal{D}}(\boldsymbol{\theta}', \boldsymbol{\theta})$ such that

$$\kappa(\boldsymbol{\theta}) - \bar\kappa = \frac{1}{\pi} \int_{\mathcal{U}} \mathrm{d}^2\theta' \, \mathcal{R}\mathrm{e}\left[\tilde{\mathcal{D}}^*(\boldsymbol{\theta}', \boldsymbol{\theta})\gamma(\boldsymbol{\theta}')\right] \quad . \tag{2.11}$$

If we use (2.4) for $\gamma$ in the preceding equation, we see that the kernel $\tilde{\mathcal{D}}$ has to satisfy the relation

$$Q(\boldsymbol{\theta}, \boldsymbol{\vartheta}) := \frac{1}{\pi^2} \int_{\mathcal{U}} \mathrm{d}^2\theta' \, \mathcal{R}\mathrm{e}\left[\tilde{\mathcal{D}}(\boldsymbol{\theta}', \boldsymbol{\theta})\mathcal{D}(\boldsymbol{\theta}' - \boldsymbol{\vartheta})\right] = \delta(\boldsymbol{\theta} - \boldsymbol{\vartheta}) - \frac{\Theta(\boldsymbol{\vartheta})}{A} \quad , \theta \in \mathcal{U}, \vartheta \in \mathbb{R}^2 \tag{2.12}$$

where $\Theta(\boldsymbol{\vartheta}) = 1$ if $\boldsymbol{\vartheta} \in \mathcal{U}$, and zero otherwise, and $A$ is the area enclosed by $\mathcal{U}$. However, we note that (2.11) is not the only possible form of an inversion equation; a more general form would include an additional integral over the boundary of $\mathcal{U}$.

In the next two sections, we present *two different strategies* to obtain inversion equations which are exact on a finite data field. By that we mean that if $\gamma$ obeys (2.4), or, alternatively, if $g$ obeys (2.8), then the surface mass density $\kappa$ can be obtained up to a constant. However, since the value of $g$ has to be determined by averaging over observed galaxy images, the relation between (the observed) $g$ and (the true) $\kappa$ will deviate from (2.8), due to noise coming from the discreteness of galaxy images, the intrinsic ellipticity of sources etc. In Sect. 3 we will summarize and generalize the work previously done on finite-field inversion formulae which are all exact for 'perfect data', but differ from their behaviour with respect to observational 'noise'. In Sect. 4, we will then obtain a new inversion method which is motivated by the fact that a certain 'noise component' can be identified as such and can be 'filtered out' in the reconstruction. We shall then, in Sect. 5, demonstrate that our new inversion formulae is superiour to the other finite-field reconstructions published (Paper III, KSFWB, Bartelmann 1995).

---

[1] Note that $|\gamma|$ decreases at most as $1/|\boldsymbol{\theta}|^2$, and for an isothermal distribution, $\gamma \propto 1/|\boldsymbol{\theta}|$.

[2] It is important to note that the shear $\gamma$ is *not* an observable in general; only for the case of weak lensing, the observable $g$ – see Eq. (2.8) – can be linearized, and in this linear approximation, $\gamma$ becomes the observable.

## 3 Finite field kernels: Averaging over line integrals

The first inversion formula on a finite field was derived in Paper III; its derivation started from the equation

$$\nabla \kappa(\boldsymbol{\theta}) = -\begin{pmatrix} \gamma_{1,1} + \gamma_{2,2} \\ \gamma_{2,1} - \gamma_{1,2} \end{pmatrix} \equiv \mathbf{U}(\boldsymbol{\theta}) \quad , \tag{3.1}$$

which was derived by Kaiser (1995) by combining appropriate combinations of third derivatives of the deflection potential $\psi$ and using (2.3). Its validity can also be checked by evaluating the gradient of (2.6b) and using that $\Delta_{\boldsymbol{\theta}'}\left(-\frac{1}{2\pi}\ln\left|\boldsymbol{\theta}-\boldsymbol{\theta}'\right|\right) = \delta(\boldsymbol{\theta}'-\boldsymbol{\theta})$. Using the definition (2.8) of $g$, Kaiser (1995) also obtained the relation

$$\nabla K(\boldsymbol{\theta}) = \frac{1}{1-g_1^2-g_2^2}\begin{pmatrix} 1+g_1 & g_2 \\ g_2 & 1-g_1 \end{pmatrix}\begin{pmatrix} g_{1,1}+g_{2,2} \\ g_{2,1}-g_{1,2} \end{pmatrix} \equiv \mathbf{u}(\boldsymbol{\theta}) \quad , \tag{3.2}$$

where

$$K(\boldsymbol{\theta}) := \ln(1-\kappa(\boldsymbol{\theta})) \quad . \tag{3.3}$$

These relations show that if $\gamma$ is assumed to be known, $\kappa$ can be determined only up to an additive constant. However, only for weak lenses ($\kappa \ll 1$) one can use the approximation $\gamma \approx g$. From $g$, $K$ can be determined up to an additive constant, or $(1-\kappa)$ can be determined only up to a multiplicative constant.

In Paper III, (3.1) was solved by integrating $\nabla \kappa$ over curves $\mathbf{l}_\lambda(t,\boldsymbol{\theta})$ connecting a point $\mathbf{b}(\lambda)$, $0 \le \lambda \le \Lambda$, of the boundary of $\mathcal{U}$ with the point $\boldsymbol{\theta}$, and then averaging over all points of the boundary,

$$\kappa(\boldsymbol{\theta}) = \frac{1}{\Lambda}\int_0^\Lambda \mathrm{d}\lambda \int_0^1 \mathrm{d}t\, \frac{\mathrm{d}\mathbf{l}_\lambda(t;\boldsymbol{\theta})}{\mathrm{d}t}\cdot\mathbf{U}(\mathbf{l}_\lambda(t;\boldsymbol{\theta})) + \frac{1}{\Lambda}\int_0^\Lambda \mathrm{d}\lambda\, \kappa(\mathbf{b}(\lambda)) \quad . \tag{3.4}$$

Since the second term is independent of $\boldsymbol{\theta}$, this equation yields $\kappa(\boldsymbol{\theta})$ up to an additive constant (i.e., the mean of $\kappa$ on the boundary of $\mathcal{U}$). Still there is much freedom left in the choice of the curves $\mathbf{l}_\lambda(t,\boldsymbol{\theta})$. For a particular choice of these curves, results were presented in Paper III. If the curves $\mathbf{l}_\lambda(t, \boldsymbol{\theta})$ for fixed $\boldsymbol{\theta}$ are chosen such that they do not intersect and that they cover the whole region $\mathcal{U}$, then there is a one-to-one relation between $(t, \lambda)$ and a point $\boldsymbol{\theta}' = \mathbf{l}_\lambda(t, \boldsymbol{\theta})$. In this case, we can rewrite (3.4) as

$$\kappa(\boldsymbol{\theta}) = \frac{1}{\Lambda} \int_\mathcal{U} \mathrm{d}^2\theta' \left| \frac{\mathrm{d}\mathbf{l}_\lambda(t, \boldsymbol{\theta})}{\mathrm{d}t} \times \frac{\mathrm{d}\mathbf{l}_\lambda(t, \boldsymbol{\theta})}{\mathrm{d}\lambda} \right|^{-1} \frac{\mathrm{d}\mathbf{l}_\lambda(t; \boldsymbol{\theta})}{\mathrm{d}t} \cdot \mathbf{U}(\boldsymbol{\theta}') + \frac{1}{\Lambda} \int_0^\Lambda \mathrm{d}\lambda\, \kappa(\mathbf{b}(\lambda)) \quad , \quad (3.5)$$

where we have defined for two 2-dimensional vectors $\mathbf{a}$ and $\mathbf{b}$,

$$\mathbf{a} \times \mathbf{b} = a_1 b_2 - a_2 b_1 \quad . \tag{3.6}$$

In Paper III the curves $\mathbf{l}$ where chosen in such a way that the integrand in (3.5) agrees locally with that of (2.6b), which led to the fairly artificially looking integration contours.

By generalizing the preceding method, KSFWB have noted that the starting points of the curves $\mathbf{l}$ need not be confined to the boundary of $\mathcal{U}$. One can proceed as follows: define curves $\mathbf{l}(t; \boldsymbol{\theta}, \boldsymbol{\theta}_0)$ with $\mathbf{l}(0; \boldsymbol{\theta}, \boldsymbol{\theta}_0) = \boldsymbol{\theta}_0$, $\mathbf{l}(1; \boldsymbol{\theta}, \boldsymbol{\theta}_0) = \boldsymbol{\theta}$, i.e., which connect the points $\boldsymbol{\theta}_0$ and $\boldsymbol{\theta}$. One can then integrate the differential equation (3.1) along each such curve. Averaging over all starting points, with an arbitrary weight function $w(\boldsymbol{\theta}_0)$, one obtains

$$\kappa(\boldsymbol{\theta}) = \frac{1}{\int_\mathcal{U} \mathrm{d}^2\theta_0\, w(\boldsymbol{\theta}_0)} \left[ \int_\mathcal{U} \mathrm{d}^2\theta_0\, w(\boldsymbol{\theta}_0) \int_0^1 \mathrm{d}t \frac{\mathrm{d}\mathbf{l}(t; \boldsymbol{\theta}, \boldsymbol{\theta}_0)}{\mathrm{d}t} \cdot \mathbf{U}\left(\mathbf{l}(t; \boldsymbol{\theta}, \boldsymbol{\theta}_0)\right) \right.$$
$$\left. + \int_\mathcal{U} \mathrm{d}^2\theta_0\, w(\boldsymbol{\theta}_0)\, \kappa(\boldsymbol{\theta}_0) \right] \quad . \tag{3.7}$$

KSFWB have considered the special cases that $w \equiv 1$ and that the curves $\mathbf{l}(t; \boldsymbol{\theta}, \boldsymbol{\theta}_0)$ are straight lines (which implicitly assumes that the region $\mathcal{U}$ is convex). In this case, the constant, represented by the second term in (3.7), becomes the mean surface mass density $\bar{\kappa}$ over the region $\mathcal{U}$. Bartelmann (1995) has given an explicit equation for the resulting inversion equation,

$$\kappa(\boldsymbol{\theta}) - \bar{\kappa} = \int_\mathcal{U} \mathrm{d}^2\theta'\, \mathbf{H}^\mathrm{B}(\boldsymbol{\theta}', \boldsymbol{\theta}) \cdot \mathbf{U}(\boldsymbol{\theta}') \quad , \tag{3.8}$$

with

$$\mathbf{H}^\mathrm{B}(\boldsymbol{\theta}', \boldsymbol{\theta}) = \frac{1}{2A} \left( \frac{R^2(\boldsymbol{\theta}', \boldsymbol{\theta})}{|\boldsymbol{\theta} - \boldsymbol{\theta}'|^2} - 1 \right) (\boldsymbol{\theta} - \boldsymbol{\theta}') \quad , \tag{3.9}$$

where $R(\boldsymbol{\theta}', \boldsymbol{\theta})$ is the length of a line segment from $\boldsymbol{\theta}$ to the boundary of $\mathcal{U}$ passing through $\boldsymbol{\theta}'$.

All the preceding equations are also valid if we replace $\kappa(\boldsymbol{\theta})$ by $K(\boldsymbol{\theta})$ – see Eq. (3.3) – and $\mathbf{U}(\boldsymbol{\theta})$ by $\mathbf{u}(\boldsymbol{\theta})$. In order to apply these inversion formulae, one needs to construct a continuous, differentiable (complex) function $g(\boldsymbol{\theta})$ from observed galaxy images, so that $\mathbf{u}(\boldsymbol{\theta})$ can be obtained from differentiation. In Sect. 5 we shall describe a simple method for obtaining such a smooth function. We also note that Eqs. (3.5) and (3.8) can be integrated by parts to remove the derivatives from $\gamma$; in the case of (3.8), this yields (2.11), with

$$\tilde{\mathcal{D}}^{\mathrm{B}}(\boldsymbol{\theta}',\boldsymbol{\theta}) = \pi\left[\left(\frac{\partial H_1^{\mathrm{B}}}{\partial \theta_1'} - \frac{\partial H_2^{\mathrm{B}}}{\partial \theta_2'}\right) + \mathrm{i}\left(\frac{\partial H_1^{\mathrm{B}}}{\partial \theta_2'} + \frac{\partial H_2^{\mathrm{B}}}{\partial \theta_1'}\right)\right] \quad, \tag{3.10}$$

due to the fact that $\mathbf{H}^{\mathrm{B}}(\boldsymbol{\theta}',\boldsymbol{\theta})$ vanishes for $\boldsymbol{\theta}'$ on the boundary of $\mathcal{U}$. For a more general kernel vector $\mathbf{H}$ which does not vanish on the boundary of $\mathcal{U}$, this partial integration will yield boundary terms. Owing to the nonlinearity of $\mathbf{u}$ in $g$, this partial integration cannot be carried out in the equations for $K$.

We can now easily check that $\tilde{\mathcal{D}}$ in (3.10) satisfies the constraint equation (2.12). Inserting (3.10) into (2.12) yields [with $\mathbf{H}^{\mathrm{B}} \equiv \mathbf{H}^{\mathrm{B}}(\boldsymbol{\theta}',\boldsymbol{\theta})$]

$$Q(\boldsymbol{\theta},\boldsymbol{\vartheta}) = \frac{1}{\pi}\int_{\mathcal{U}} \mathrm{d}^2\theta' \left[\left(H_{1,1}^{\mathrm{B}} - H_{2,2}^{\mathrm{B}}\right)\mathcal{D}_1(\boldsymbol{\theta}' - \boldsymbol{\vartheta}) + \left(H_{1,2}^{\mathrm{B}} + H_{2,1}^{\mathrm{B}}\right)\mathcal{D}_2(\boldsymbol{\theta}' - \boldsymbol{\vartheta})\right]$$
$$= \frac{-1}{\pi}\int_{\mathcal{U}} \mathrm{d}^2\theta' \left[H_1^{\mathrm{B}}\left(\mathcal{D}_{1,1} + \mathcal{D}_{2,2}\right) + H_2^{\mathrm{B}}\left(\mathcal{D}_{2,1} - \mathcal{D}_{1,2}\right)\right] \quad,$$

where we used again that $\mathbf{H}^{\mathrm{B}}$ vanishes on the boundary of $\mathcal{U}$, and all indices separated by commas denote partial differentiation with respect to $\theta_i'$. We then note that $\mathcal{D}$ can be written as second partial derivatives of the function $f(\boldsymbol{\theta}) := \ln|\boldsymbol{\theta}|$, as

$$\mathcal{D}_1(\boldsymbol{\theta}) = \frac{f_{,22}(\boldsymbol{\theta}) - f_{,11}(\boldsymbol{\theta})}{2} \quad, \quad \mathcal{D}_2(\boldsymbol{\theta}) = -f_{,12}(\boldsymbol{\theta}) \quad,$$

from which it follows that

$$\mathcal{D}_{1,1}(\boldsymbol{\theta}) + \mathcal{D}_{2,2}(\boldsymbol{\theta}) = -\frac{1}{2}\left(f_{,111}(\boldsymbol{\theta}) + f_{,122}(\boldsymbol{\theta})\right) = -\frac{1}{2}(\Delta f)_{,1} = -\pi\frac{\partial}{\partial \theta_1}\delta(\boldsymbol{\theta}) \quad,$$
$$\mathcal{D}_{2,1}(\boldsymbol{\theta}) - \mathcal{D}_{1,2}(\boldsymbol{\theta}) = -\frac{1}{2}\left(f_{,112}(\boldsymbol{\theta}) + f_{,222}(\boldsymbol{\theta})\right) = -\frac{1}{2}(\Delta f)_{,2} = -\pi\frac{\partial}{\partial \theta_2}\delta(\boldsymbol{\theta}) \quad,$$

so that, after another partial integration, we obtain

$$Q(\boldsymbol{\theta},\boldsymbol{\vartheta}) = -\left(\frac{\partial H_1^{\mathrm{B}}(\boldsymbol{\vartheta},\boldsymbol{\theta})}{\partial \vartheta_1} + \frac{\partial H_2^{\mathrm{B}}(\boldsymbol{\vartheta},\boldsymbol{\theta})}{\partial \vartheta_2}\right)\Theta(\boldsymbol{\vartheta}) \quad. \tag{3.11}$$

Using the differentiation of $\mathbf{H}^{\mathrm{B}}$ as described in Bartelmann (1995), we finally find that

$$Q(\boldsymbol{\theta},\boldsymbol{\vartheta}) = -\frac{\Theta(\boldsymbol{\vartheta})}{A} \quad \text{for } \boldsymbol{\vartheta} \neq \boldsymbol{\theta} \quad. \tag{3.12}$$

On the other hand,

$$\int_{\mathcal{U}} \mathrm{d}^2\vartheta\, Q(\boldsymbol{\theta},\boldsymbol{\vartheta}) = -\int_{\mathcal{U}} \mathrm{d}^2\vartheta\, \nabla\cdot\mathbf{H}^{\mathrm{B}}(\boldsymbol{\vartheta},\boldsymbol{\theta}) = 0 \quad, \tag{3.13}$$

again because $\mathbf{H}$ vanishes on the boundary. Combining (3.12 & 3.13), we now see the validity of (2.12).

Using Eq. (3.7) to derive an inversion formula both has its strength and weakness, both essentially for the same reason: there is so much freedom left in the choice of the weight function $w(\boldsymbol{\theta}_0)$ and the integration curves $\mathbf{l}(t;\boldsymbol{\theta},\boldsymbol{\theta}_0)$ that it will be difficult to single out 'the best' choice for a given problem. However, this large freedom of choice may be used profitably, e.g., if the data set has 'holes' (which can be generated, for example, by some bright cluster galaxies which do not allow to measure $g$ in a certain region of the cluster) to avoid such regions. We shall not discuss this issue further in this paper, but assume that the data for $g$ are available on the entire region $\mathcal{U}$.

# 4 Finite field kernels: Noise filtering

In this section we derive another inversion equation, again starting from (3.2) [or (3.1); we shall derive the inversion equation from (3.2), but can afterwards replace $K$ by $\kappa$ and $\mathbf{u}$ by $\mathbf{U}$ to obtain an inversion equation of (3.1)]. The starting point of our considerations is the simple observation that the vector field $\mathbf{u}(\boldsymbol{\theta})$, obtained from observed distorted galaxy images, will in general not be a gradient field, but due to the noise, caused by the discreteness of galaxy images and their intrinsic ellipticity, it will also contain a rotational component. *If $\mathbf{u}(\boldsymbol{\theta})$ is not a gradient field, equation (3.2) does not have a solution $K(\boldsymbol{\theta})$ ! On the other hand, if $\mathbf{u}(\boldsymbol{\theta})$ is a gradient field, then all finite-field inversion equations are equivalent!* [Of course, they may differ in their numerical applications; for example, the function (3.10) is not continuous if the region $\mathcal{U}$ has corners and must therefore be applied numerically with great care to avoid artificial features in the resulting $\kappa$-distribution]. If $\mathbf{u}$ is not a gradient field, the different finite-field inversion equations *differ in their treatment of the rotational part of $\mathbf{u}$*. In other words, in this case the result of equations like (3.4) and (3.7) depends on the choice of the integration paths $\mathbf{l}$.

Let us decompose the field $\mathbf{u}(\boldsymbol{\theta})$ into a gradient and a rotational part,

$$\mathbf{u}(\boldsymbol{\theta}) = \nabla \tilde{K}(\boldsymbol{\theta}) + \mathbf{rot}\, s(\boldsymbol{\theta}) \equiv \nabla \tilde{K}(\boldsymbol{\theta}) + \begin{pmatrix} \frac{\partial s}{\partial \theta_2} \\ -\frac{\partial s}{\partial \theta_1} \end{pmatrix} \quad , \tag{4.1}$$

where $s(\boldsymbol{\theta})$ is a scalar field, and where we have defined the 'rotation' of a scalar in the second step, which is the gradient of the scalar field rotated by $\pi/2$. Unfortunately, this decomposition is *not* unique, as shown by the simple example

$$\nabla(\theta_1^2) = \begin{pmatrix} 2\theta_1 \\ 0 \end{pmatrix} = \nabla\left(|\boldsymbol{\theta}|^2/2\right) + \mathbf{rot}(\theta_1\theta_2) \quad .$$

In order to determine $\nabla \tilde{K}$ from $\mathbf{u}$, we thus need additional specifications for the decomposition. Since the rotational part of $\mathbf{u}$ is due to noise, it appears natural to perform the decomposition such that the rotational part is in some sense 'minimized'. In particular, we require that $\mathbf{rot}\,s(\boldsymbol{\theta})$ should vanish if $\mathbf{u}$ is a gradient field. Also, we require that there is no systematic rotational component over $\mathcal{U}$, i.e., that the vector $\mathbf{rot}\,s$, averaged within $\mathcal{U}$, should vanish. With these two conditions, the decomposition (4.1) is now uniquely determined: from differentiation of (4.1), we obtain

$$\frac{\partial u_1}{\partial \theta_2} - \frac{\partial u_2}{\partial \theta_1} \equiv \nabla \times \mathbf{u} = \Delta s \quad , \tag{4.2a}$$

where $\Delta$ is the Laplace operator. If we require that

$$s = 0 \quad \text{on the boundary } \partial \mathcal{U} \text{ of } \mathcal{U}, \tag{4.2b}$$

then both of the above requirements are satisfied, and $\mathbf{rot}\, s$ is uniquely determined. We could of course choose any constant in (4.2b) without changing $\mathbf{rot}\, s$. In fact, we shall not need to solve the boundary-value-problem (4.2), but only make use of the existence of its solution.[3] Then, we identify $\nabla \tilde{K}$ with $\nabla K$; although these two vector fields will be different in reality, since also the gradient part of $\mathbf{u}$ includes noise, we cannot separate this noise component from a true signal. The above identification is motivated by the expectation that the difference between $\nabla \tilde{K}$ and $\nabla K$ will be random, with zero mean over the field $\mathcal{U}$, and spatially uncorrelated on scales larger than the smoothing scale introduced later (see Sect. 5). Noting that $K(\boldsymbol{\theta})$ can be determined only up to an additive constant, which we choose to be the average $\bar{K}$ of $K$ over $\mathcal{U}$, and that $K$ and $\mathbf{u}$ are linearly related, we can make the ansatz

$$K(\boldsymbol{\theta}) - \bar{K} = \int_{\mathcal{U}} \mathrm{d}^2 \theta' \, \mathbf{H}(\boldsymbol{\theta}', \boldsymbol{\theta}) \cdot \mathbf{u}(\boldsymbol{\theta}') \quad . \tag{4.3}$$

We now replace $\mathbf{u}(\boldsymbol{\theta}')$ by its decomposition (4.1) and integrate by parts; this yields

$$\begin{aligned} K(\boldsymbol{\theta}) - \bar{K} = &\oint_{\partial \mathcal{U}} \mathrm{d}\lambda \, K(\boldsymbol{\theta}') \, \mathbf{n}(\boldsymbol{\theta}') \cdot \mathbf{H}(\boldsymbol{\theta}', \boldsymbol{\theta}) + \oint_{\partial \mathcal{U}} \mathrm{d}\boldsymbol{\lambda} \cdot \mathbf{H}(\boldsymbol{\theta}', \boldsymbol{\theta}) \, s(\boldsymbol{\theta}') \\ &- \int_{\mathcal{U}} \mathrm{d}^2 \theta' \, K(\boldsymbol{\theta}') \, \nabla \cdot \mathbf{H}(\boldsymbol{\theta}', \boldsymbol{\theta}) - \int_{\mathcal{U}} \mathrm{d}^2 \theta' \, s(\boldsymbol{\theta}') \, \nabla \times \mathbf{H}(\boldsymbol{\theta}', \boldsymbol{\theta}) \quad , \end{aligned} \tag{4.4}$$

where $\mathrm{d}\boldsymbol{\lambda}$ is the length element on $\partial \mathcal{U}$, and all differential operators operate on $\boldsymbol{\theta}'$. Here, $\mathbf{n}$ is the outward directed normal at the boundary of $\mathcal{U}$ over which the first two integrals are taken. We now consider the four terms in turn: the second term vanishes due to (4.2b). The last term can be made to vanish if we require $\mathbf{H}(\boldsymbol{\theta}', \boldsymbol{\theta})$ to be a gradient vector field with respect to $\boldsymbol{\theta}'$,

$$\mathbf{H}(\boldsymbol{\theta}', \boldsymbol{\theta}) = \nabla \mathcal{L}(\boldsymbol{\theta}', \boldsymbol{\theta}) \quad . \tag{4.5}$$

The first term can be made to vanish if we require

$$\mathbf{H}(\boldsymbol{\theta}', \boldsymbol{\theta}) \cdot \mathbf{n}(\boldsymbol{\theta}') = 0 \quad \text{on the boundary of } \mathcal{U}. \tag{4.6}$$

---

[3] It would be possible, of course, to solve (4.2) for $s$ and to convolve the 'cleaned' data $\mathbf{u} - \mathbf{rot}\, s$ with any finite-field kernel; this would yield the same resulting mass distribution as our new inversion formula derived below. However, one then has to repeat this 'cleaning procedure' for every new data set. Instead, it is our goal to develop a kernel which filters out automatically the uniquely specified rotational component and simultaneously yields a reconstructed density field. One then has to calculate the inversion kernel only once for a given geometry of the data field.

Finally, the third term can be made to agree with the left hand side of (4.4) if we require

$$\Delta \mathcal{L}(\boldsymbol{\theta}', \boldsymbol{\theta}) = \nabla \cdot \mathbf{H}(\boldsymbol{\theta}', \boldsymbol{\theta}) = -\delta(\boldsymbol{\theta}' - \boldsymbol{\theta}) + \frac{1}{A} \quad , \tag{4.7}$$

where again $A$ is the area of the region $\mathcal{U}$. Eqs.(4.6) and (4.7) together define a Neumann boundary problem, which has a unique solution for $\mathcal{L}(\boldsymbol{\theta}', \boldsymbol{\theta})$, up to an additive constant, i.e., it has a unique solution for $\mathbf{H}(\boldsymbol{\theta}', \boldsymbol{\theta})$. In the appendix, we present a closed-form expression for $\mathbf{H}$ for a circle, and give a fairly detailed description for the calculation of $\mathbf{H}$ for a rectangle.

Finally, we want to point out that the vector field $\mathbf{H}$ needs to be calculated only once if similar field geometries are considered, due to the following scaling relation (which can be easily verified, using the explicit equations in the appendix): if $L$ denotes a scale of the data field (e.g., the side length of a rectangle or the radius of a circle), we introduce scaled coordinates $\mathbf{x}$ as $\boldsymbol{\theta} = L\mathbf{x}$, $\boldsymbol{\theta}' = L\mathbf{x}'$. Then,

$$\mathbf{H}(\boldsymbol{\theta}', \boldsymbol{\theta}) = \frac{1}{L}\hat{\mathbf{H}}(\mathbf{x}', \mathbf{x}) \quad , \tag{4.8}$$

where $\hat{\mathbf{H}}$ is the kernel for $L = 1$.

## 5 Comparison of inversion methods

In this section, we apply the various inversion techniques discussed in this paper and in Paper III to synthetic data. We choose a lensing mass distribution, distribute sources randomly in position, and assign an ellipticity to them according to an assumed intrinsic ellipticity distribution. For each such source, we then calculate the 'observed' ellipticity from the intrinsic one and using the lens model. The data field to be analyzed thus consists of galaxy positions and ellipticities. Applying the four inversion techniques yields four reconstructed $K$-distributions, which are compared to the true field, $K_{\text{true}} = \ln(1 - \kappa_{\text{true}})$. The difference $K - K_{\text{true}}$ is decomposed into its Fourier modes. We repeat this analysis for different source distributions many ($\approx 50$) times to obtain the power spectrum of the reconstruction-error field. Throughout this section, all angles are measured in arcminutes. The data field is assumed to be a square of side length $L$.

### 5.1 Method

#### 5.1.1 Lens models
The lensing mass distributions which we use here are constructed from superpositions of simple components of the form

$$\kappa(\boldsymbol{\theta}) = \kappa_0 \frac{1 + \left(|\boldsymbol{\theta} - \boldsymbol{\theta}_0|^2 / 2\theta_c^2\right)}{\left[1 + \left(|\boldsymbol{\theta} - \boldsymbol{\theta}_0|^2 / \theta_c^2\right)\right]^{3/2}} \quad , \tag{5.1}$$

where $\kappa_0$ is the central surface mass density, $\boldsymbol{\theta}_0$ the center of the mass component, and $\theta_c$ its core radius. Note that for $|\boldsymbol{\theta}| \gg \theta_c$, the density behaves like an isothermal sphere. The Einstein radius of the corresponding singular isothermal sphere (i.e., that with the same behaviour for $|\boldsymbol{\theta}| \gg \theta_c$) is $\theta_E = \kappa_0 \theta_c$. Four different mass distributions are investigated: Lens B is a strong, but sub-critical lens mass distribution consisting of two components of the form (5.1), e.g. a bimodal cluster or a cluster and a group of galaxies along the line of sight, and its center of mass is near the center of the data field; lens A is simular to B, but displaced to the boundary of the field; lens C is a weakly lensing mass distribution consisting of 3 components (e.g., groups of galaxies, or poor clusters) of the nearly isothermal mass distribution (5.1); two of the components are centered near the middle of the data field, one is at its left edge; lens D is the same as C but the third mass distribution at the left edge is removed. For each lens model the positions of the components, their core radii and their central surface mass densities are displayed in Table 1. To relate their lensing strength to that of a SIS-model we have added the corresponding velocity dipsersion of the SIS-lens in the table. For all models a surface- and contour plot of the field $-K(\mathbf{x}) = -\ln(1 - \kappa(\mathbf{x}))$ is shown in Fig. 1.

**Table 1.** Parameters of the four lenses A, B, C and D: Lens A, B and D are superpositions of two ($\alpha$ and $\beta$) components of the form (5.1), lens C consists of three components ($\alpha$, $\beta$ and $\gamma$). The first three lines contain the peak positions $\boldsymbol{\theta}_0$ of the mass components, where the lower left corner of the field is the origin of reference, and all length scales are in units of arcminutes; the lines 4 to 6 (7 to 9) contain the core radii (central mass densities) of the superposed mass distributions; in the last three lines the velocity dispersions of the corresponding isothermal sphere models are indicated.

| Parameter | LensA | LensB | LensC | LensD |
|---|---|---|---|---|
| $\theta_{0\alpha}$ | (6.5, 3.5) | (3.0, 5.0) | (3.0, 5.0) | (3.0, 5.0) |
| $\theta_{0\beta}$ | (5.5, 5.5) | (5.0, 3.0) | (5.0, 3.0) | (5.0, 3.0) |
| $\theta_{0\gamma}$ | — | — | (0.0, 3.5) | — |
| $\theta_{c\alpha}$ | 2.0 | 2.0 | 0.9 | 0.9 |
| $\theta_{c\beta}$ | 1.0 | 1.0 | 0.9. | 0.9 |
| $\theta_{c\gamma}$ | — | — | 2.0 | — |
| $\kappa_{0\alpha}$ | 0.5 | 0.5 | 0.3 | 0.3 |
| $\kappa_{0\beta}$ | 0.3 | 0.3 | 0.2 | 0.2 |
| $\kappa_{0\gamma}$ | — | — | 0.2 | — |
| $\sigma_\alpha$ | 1400 | 1400 | 735 | 735 |
| $\sigma_\beta$ | 775 | 775 | 600 | 600 |
| $\sigma_\gamma$ | — | — | 895 | — |

**Fig. 1.** Contour- and surface plot for the function $-K(\mathbf{x}) = -\ln[1 - \kappa(\mathbf{x})]$, where $\kappa$ is the two-dimensional surface mass density of the lens A,B,C and D respectively. In each panel, the size of the field is $7.'5 \times 7.'5$, and $K$ is calculated on a $50 \times 50$ grid; the spacing in the contour lines is 0.1 for lens A and lens B and 0.05 for lens C and D

### 5.1.2 Generation of the source data and calculation of the 'observed' data field

We distribute sources randomly in $\boldsymbol{\theta}$ (i.e. on the observer and not on the source sky), which is a valid procedure due to the fact that the magnification bias basically vanishes for the faint blue galaxy population, since their source counts behave nearly as $N(>S) \propto S^{-1}$. Their density is $n_0/(\text{arcmin})^2$, and typically $n_0 = 50$. The intrinsic ellipticity $\epsilon^{(s)}$ of the sources is defined as in Paper III. [Briefly, $\epsilon^{(s)}$ is a complex number whose modulus is $(1-r)/(1+r)$, for an elliptical source with axial ratio $r \leq 1$; for general brightness profiles of the sources, $\epsilon^{(s)}$ is defined in terms of the tensor of second brightness moments. The phase of $\epsilon^{(s)}$ describes the orientation of the source.] In our simulations, the intrinsic ellipticity $\epsilon^{(s)}$ was drawn from a Gaussian distribution of the form

$$p_{\mathrm{s}}(\epsilon^{(s)}) = \frac{1}{\pi \rho^2 (1 - \mathrm{e}^{-1/\rho^2})} \mathrm{e}^{-|\epsilon^{(s)}|^2/\rho^2} \quad , \tag{5.2a}$$

where $p_{\mathrm{s}}(\epsilon^{(s)}) \, \mathrm{d}^2 \epsilon^{(s)}$ is the probability that the source ellipticity lies within $\mathrm{d}^2 \epsilon^{(s)}$ of $\epsilon^{(s)}$. Hence, the quantity $\rho$ controls the width of the intrinsic ellipticity distribution, and we expect that with increasing $\rho$, the reconstructed mass density will become noisier.

Other authors have used ellipticity distributions different from (5.2a); in Papers I&II, and in KS, Gaussian distributions in the absolute value of

$$\chi := \frac{2\epsilon}{1 + |\epsilon|^2} \quad , \tag{5.2b}$$

were used (see Paper II, eq. (2.22) and KS, page 445, second column, and note that the quantity $|e|$ used by KS is equal to $|\chi|$ defined in Paper II). Hence, these authors used an ellipticity distribution of the following form:

$$p_{\mathrm{s}}(\chi^{(s)}) = \frac{1}{\pi R^2 (1 - \mathrm{e}^{-1/R^2})} \mathrm{e}^{-|\chi^{(s)}|^2/R^2} \quad . \tag{5.2c}$$

Using that for sources with elliptical isophotes $|\epsilon^{(s)}| = \frac{1-r}{1+r}$ and $|\chi^{(s)}| = \frac{1-r^2}{1+r^2}$, we have calculated the probabilities $p_\rho(r)$ and $p_R(r)$ that a source drawn from an ellipticity distribution (5.2a) or (5.2c) has an axis ratio $r$; the widths $R$ and $\rho$ were chosen as $R, \rho \in [0.1, 0.2, 0.3, 0.4, 0.5]$. The results can be seen in Fig. 2: the solid curves, $p_\rho(r)$, were obtained from (5.2a), where with increasing $\rho$ the maximum is shifted to the left, i.e., elliptical sources become more likely; the analogous curves, $p_R(r)$ for (5.2c) are dashed. One can see from Fig. 2 that for $\rho \lesssim 0.2$, $p_\rho(r) \approx p_{R=2\rho}(r)$; this is due to the fact that for small $\epsilon$, (5.2b) becomes $\chi \approx 2\epsilon$. In particular, this implies that the simulations in KS and in Papers I&II, where an ellipticity distribution of the form (5.2c) with $R = 0.3$ was used, consider only rather circular sources; this would correspond to a width of $\rho = 0.15$ in our simulations.

Finally, the 'observed' ellipticity for each galaxy is calculated from the lens model, using

$$\epsilon = \frac{\epsilon^{(s)} - g}{1 - g^* \epsilon^{(s)}} \quad . \tag{5.3}$$

We want to stress that this transformation law is valid only for non-critical clusters, where $|g| < 1$ everywhere. In the more general case of critical clusters, one should use the ellipticity parameter $\chi$, as defined in Paper I. Thus, the synthetic data set consists of a set of values for the galaxy positions $\boldsymbol{\theta}_{gi}$ and their corresponding ellipticity $\epsilon_i$.

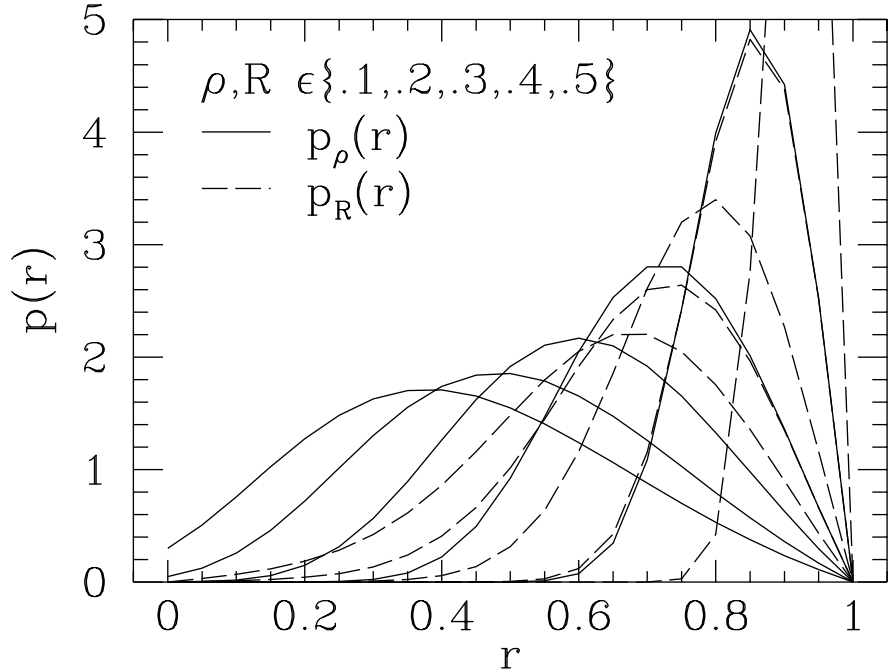

**Fig. 2.** Solid lines: probabilities $p_\rho(r)$ that a galaxy drawn from the ellipticity distribution (5.2a) has an axis ratio $r$, where $\rho \in \{0.1, 0.2, 0.3, 0.4, 0.5\}$. Dashed lines: probabilities $p_R(r)$ that a galaxy drawn from the ellipticity distribution (5.2c) has an axis ratio $r$, where $R \in \{0.1, 0.2, 0.3, 0.4, 0.5\}$. For $\rho \lesssim 0.2$, the probabilities $p_\rho(r)$ and $p_{R=2\rho}(r)$ roughly coincide.

### 5.1.3 Determination of $g$, 'inner-smoothing'

If we average the observed ellipticities of the images over a region of the lens within which $g$ is approximately constant, their expectation value is

$$\langle \epsilon \rangle = -g \quad , \tag{5.4}$$

due to the fact that the sources are intrinsically randomly oriented. This fact has been shown by Schramm & Kayser (1994) and can be obtained simply from an angular integration of (5.3). On a regular grid of points

$$\boldsymbol{\theta}'_{ij} = (ia, ja) \quad , \quad 0 \le i, j \le N \quad , \tag{5.5}$$

with $a = L/N$ being the size of a gridcell, and $N$ the number of gridpoints per dimension on the square, we have calculated an estimate of $g(\boldsymbol{\theta}'_{ij}) \equiv g_{ij}$ by a weighted average, according to (5.4),

$$g_{ij} = -\frac{\sum_m w_m \epsilon_m}{\sum_m w_m} \quad , \tag{5.6}$$

with Gaussian weights

$$w_m \propto \exp\left(\frac{-(\boldsymbol{\theta}'_{ij} - \boldsymbol{\theta}_{gm})^2}{\Delta\theta_{ij}^2}\right) \quad , \tag{5.7}$$

and the smoothing scale $\Delta\theta$ can be chosen appropriately. As described in detail in Paper II, it is useful to adapt the smoothing scale to the strength of the signal; i.e., in a region where $|g|$ is large, the ellipticity of the observed galaxy images will be dominated by the shear effect of the lens, and an accurate value of $g$ can be obtained by averaging over a few images only, whereas in regions of weaker shear, the average should extend over a larger number of images. We take the same smoothing procedure as in Paper II, i.e., we choose

$$\Delta\theta_{ij} = \Delta\theta_0 \left(1 - |\delta_{i,j-1}|^2\right)^\beta \quad , \tag{5.8}$$

where the prefactor $\Delta\theta_0$ should be of the order of several times the mean separation of galaxy images (e.g., $\Delta\theta_0 \approx \frac{8}{\sqrt{n_0}}\rho$), and the exponent $\beta$ determines how the smoothing length decreases with increasing $|g|$, i.e., increasing $|\delta|$. For the inversions shown here, we chose $\beta = 2$. Since we later will also smooth the reconstructed density field, we call the smoothing of the ellipticities to get an estimate of $g$ the 'inner smoothing' and that of the density field 'outer smoothing'. We point out that we applied the same smoothing procedure (with the same smoothing lengths) to all four inversion methods. Changing the smoothing lengths affects the quality of the reconstruction differently for each method, and each method will have a different optimal smoothing length with respect to a given quality criterium for the reconstruction. For instance, if the quality criterium is to resolve the heights of density peaks, then the maximally acceptable smoothing length of the KS inversion will be smaller than that of the other three inversion techniques. However, we want to emphasize that we do not develop an optimized inversion with respect to smoothing for each inversion kernel in this Paper, but we compare the noise sensitivities and systematics for the four kernels; in particular we show that the kernel developed in Sect. 4 is the least noise-sensitive finite-field kernel, and the increase in noise compared to the KS reconstruction is compensated by far by the loss of the boundary artefacts.

### 5.2 Discretized form of the inversion formulae

#### 5.2.1 KS-inversion

With the values of $g_{ij}$ obtained according to (5.6), we calculate the KS estimate [see Eq. (2.10a)] of $\kappa$ iteratively, as suggested in Paper II

$$\kappa^{n+1}(\boldsymbol{\theta}'_{kl}) = \frac{a^2}{\pi} \sum_{i,j=0}^{N} W_i W_j \left[1 - \kappa^n(\boldsymbol{\theta}'_{ij})\right] \mathcal{R}e\left[\mathcal{D}^* \left(\boldsymbol{\theta}'_{kl} - \boldsymbol{\theta}'_{ij}\right) g_{ij}\right] \quad , \quad n \geq 0 \quad . \tag{5.9}$$

The starting value of this iteration is $\kappa^0 \equiv 0$, and convergence is obtained after only a few iterations. Here, $W_i = 1/2$ for $i = 0$ and for $i = N$, and $W_i = 1$ otherwise. The resulting solution is then interpolated on the grid

$$\boldsymbol{\theta}_{kl} = (\,[k - 1/2]\,a\,,\,[l - 1/2]\,a\,) \tag{5.10}$$

by

$$\kappa_{kl} \equiv \kappa(\boldsymbol{\theta}_{kl}) = \frac{1}{4}\left[\kappa(\boldsymbol{\theta}'_{kl}) + \kappa(\boldsymbol{\theta}'_{k-1,l}) + \kappa(\boldsymbol{\theta}'_{k,l-1}) + \kappa(\boldsymbol{\theta}'_{k-1,l-1})\right] \quad . \tag{5.11}$$

Since we use the nonlinear form of the reconstruction, i.e., we take $g$ as the observable, not the shear $\gamma$, the result of the iteration determines $K$, as defined in (3.3), up to an additive constant. Hence, our estimate of $K_{kl} \equiv K(\boldsymbol{\theta}_{kl})$ from the KS method, uncorrected for the finite field, is

$$K_{kl}^{\mathrm{KS}} - K_0 = \ln(1 - \kappa_{kl}) \quad , \tag{5.12}$$

where the constant $K_0$ cannot be determined.

**5.2.2 Finite field kernels**

The other three reconstruction methods we want to investigate here are all based on the vector field $\mathbf{u}(\boldsymbol{\theta})$, defined in (3.2). We have calculated this vector field on the grid $\boldsymbol{\theta}'_{ij}$ from (5.6), using finite differencing, thus obtaining the values $\mathbf{u}_{ij} \equiv \mathbf{u}(\boldsymbol{\theta}'_{ij})$. Using (3.8), with $\kappa$ replaced by $K$ and $\mathbf{U}$ replaced by $\mathbf{u}$, we obtain a second estimate for $K$ from the method developed by KSFWB and Bartelmann (1995), which we shall call for notational simplicity

$$K_{kl}^{\mathrm{B}} - \bar{K} = a^2 \sum_{i,j=0}^{N} W_i W_j \mathbf{H}^{\mathrm{B}}(\boldsymbol{\theta}'_{ij}, \boldsymbol{\theta}_{kl}) \cdot \mathbf{u}_{ij} \quad , \tag{5.13}$$

with $\mathbf{H}^{\mathrm{B}}$ given in (3.9). Similarly, the estimate from our kernel developed in this paper becomes, according to (4.3),

$$K_{kl}^{\mathrm{SS}} - \bar{K} = a^2 \sum_{i,j=0}^{N} W_i W_j \mathbf{H}(\boldsymbol{\theta}'_{ij}, \boldsymbol{\theta}_{kl}) \cdot \mathbf{u}_{ij} \quad , \tag{5.14}$$

with the kernel $\mathbf{H}$ determined numerically, as described in the appendix. Finally, we obtain an estimate of $K_{kl}$ from the inversion method described in Paper III, which we shall term $K_{kl}^{\mathrm{S}}$.

**5.2.3 'Outer smoothing'**

Obviously, the smoothing introduced by $\Delta\theta$ is fairly moderate, if $\Delta\theta_0$ is of the order of the mean galaxy separation, and the resulting estimates for $K$ will be very noisy for reasonable values of the parameter $\rho$ in the ellipticity distribution (5.2). We remind the reader that we needed the first smoothing to obtain a smooth function $g(\boldsymbol{\theta})$ which can be differentiated (of course, $g$ will be fairly noisy too). Hence, we apply a second smoothing on the resulting distribution of $K$, i.e., we define for all estimates the smoothed fields

$$\tilde{K}_{kl} = \frac{\sum_{ij} K_{ij} \exp\left(-|\boldsymbol{\theta}_{ij} - \boldsymbol{\theta}_{kl}|^2/\theta_{\mathrm{sm}}^2\right)}{\sum_{ij} \exp\left(-|\boldsymbol{\theta}_{ij} - \boldsymbol{\theta}_{kl}|^2/\theta_{\mathrm{sm}}^2\right)} \quad , \tag{5.15}$$

with the same smoothing length for the KS, SS, S and B reconstructions. In principle, the smoothing length $\theta_{\mathrm{sm}}$ can depend on the position $\boldsymbol{\theta}_{kl}$, and an 'optimized' reconstruction could adjust this local smoothing scale by some appropriate measure, such as a local measure of a $\chi^2$-statistics [see Eq. (4.8) of Paper II]. We shall not develop such a scheme here. Instead, we will use $\theta_{\rm sm}$ to be a constant, and to be typically of the order of the gridsize or larger (this removes grid effects).

### 5.2.4 Power spectrum of the error-field

We then compare the resulting smoothed field $\tilde{K}$ from the various reconstruction methods with the true field $K_{\rm true} = \ln(1 - \kappa_{\rm true})$. For this purpose, we define the Fourier transform of the difference between the reconstructed field $\tilde{K}$ and the true field $K_{\rm true}$,

$$D(\mathbf{k}) = \frac{1}{L^2} \int_{\mathcal{U}} d^2\theta\, e^{i\mathbf{k}\cdot\boldsymbol{\theta}} \bigl[\tilde{K}(\boldsymbol{\theta}) - K_{\rm true}(\boldsymbol{\theta})\bigr] \quad, \tag{5.16}$$

where $\tilde{K}$ stands for $\tilde{K}^{\rm KS}$, $\tilde{K}^{\rm SS}$, $\tilde{K}^{\rm S}$ and $\tilde{K}^{\rm B}$, respectively. We then define the power spectrum of this difference to be

$$P(k) := \langle D(\mathbf{k}) D^*(\mathbf{k}) \rangle \quad, \tag{5.17}$$

where the average extends over the directions of the (appropriately binned) $\mathbf{k}$-vectors and various realizations of the source distribution.

## 5.3 Results

### 5.3.1 Results for the power spectra

For a given lens model, the power spectrum of the error field depends on several parameters: the number $N$ of gridpoints, the size $L$ of the data field, the density $n_0$ of galaxies, the width $\rho$ of the ellipticity distribution, the inner smoothing length $\theta_0$ and the outer smoothing length $\theta_{\rm sm}$. For all power spectra shown here the number $l_{\rm max}$ of simulations which have been made to calculate the power spectrum is equal to 50 and $N = 50$. If not stated otherwise, $L = 7.5$, which is the size for a currently available CCD at the CFHT (Fahlman et al. 1994, Tyson 1994). The density of galaxies varies between 40 and 80 (per square arc minute), and $0.2 \le \rho \le 0.4^4$; the inner smoothing length varies from 0.2 to 0.5 (arcminutes) and the $\theta_{\rm sm}$ is typically of the order $a/\sqrt{2}$. In order to emphasize the differences of the reconstructions at small length scales, we have plotted $P(k)k^2$ instead of $P(k)$.

We first consider reconstructions of the lens B, which is an example for a fairly strong but non-critical cluster. Since the ellipticity distribution of the faint background population is currently not well determined, we consider ellipticity distributions with $\rho$ in the range of 0.2 to 0.4. For $n_0 = 50$ and $\rho = 0.2$ we reconstructed the lens with $\theta_0 = 0.25$, $\theta_{\rm sm} = 0$ (reconstruction B1) and $\theta_0 = 0.34$, $\theta_{\rm sm} = a/\sqrt{2} = 0.11$ (reconstruction B2); the corresponding power spectra are shown in Fig. 3a and 3b.

One sees that the KS power spectrum has a large spike at $k = 2\pi/L$, which is due to the systematic artefacts near the boundary of the CCD field. Despite this, at larger wave numbers the power spectra of the KS, SS and S error look quite simular, where the quality of reconstruction is slightly decreasing from the KS- to SS- and S-method at those wave vectors; the power spectrum of the B-inversion is considerably worse, i.e., the reconstruction is rather noisy on small scales. The reconstruction B1 was obtained without any 'outer smoothing', $\theta_{sm} = 0$; the influence of an outer smoothing can be seen by comparison with Fig. 3b. In this model, the inner smoothing length was also larger, which yields a general decrease of all power spectra. The outer smoothing becomes visible for $k \gtrsim 8$ by an exponential suppression of the power spectra. For the reconstruction B2 we also show a single realization of the KS-, SS-, S- and B-inversion in Fig. 4: we have used the same source distribution for all inversions and all $K$-estimates are shifted such that the average over the field equals the average of $K^{\text{true}}$.

---
[4] Recent investigations by Brainerd et al. (1995; Brainerd 1995, private communication) indicate that the observed (seeing convolved) galaxy images down to $r \lesssim 26$ are fairly circular: the probability distribution peaks at $r = 0.88$, whereas the behaviour for small $r$ seems to be similar to the case of a probability distribution according to (5.2a) with a width of $\rho = 0.2$. Therefore, simulations with $\rho = 0.2$ seem to be quite realistic. However, since the ellipticity distribution for galaxies in the blue light is probably broader, we also present test simulations with larger values of $\rho$ below.

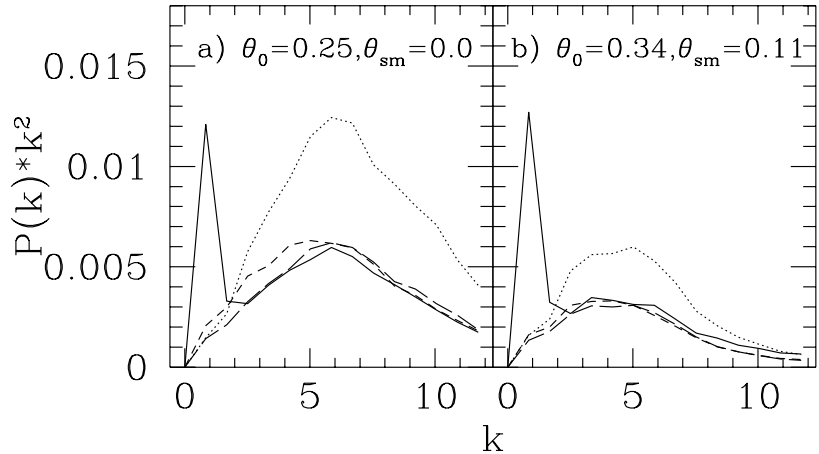

**Fig. 3.** Power spectra of the error field of the reconstructed lens B, for parameters $N = 50$, $L = 7.5$, $n_0 = 50$, $\rho = 0.2$; the inner and outer smoothing lengths were varied: (a) $\theta_0 = 0.25$, $\theta_{sm} = 0$, (b) $\theta_0 = 0.34$ and $\theta_{sm} = 0.11$. The solid (long-dashed, short-dashed and dotted) lines correspond to the KS- (SS-, S- and B-inversion) respectively. The spike in the power spectrum of the KS-inversion at $k = \frac{2\pi}{L}$ comes from the boundary artefacts of this reconstruction method; for larger values of $k$ the power spectra of the KS-, SS- and S-reconstruction roughly coincide, the B-reconstruction is fairly noisy. For wave vectors $k \geq 3$, and for all inversion techniques, the power spectrum mainly reflects 'real noise' in the inverted $K$-field; in other words, for a small smoothing length only a minor part of the power spectrum at large $k$ is due to systematic effects, such as not resolving small scale maxima of the density field due to smoothing. The exponential cutoff at $k \gtrsim 8$ in Fig. 3b is caused by the 'outer smoothing', $\theta_{sm} = 0.11$

Inspection by eye indicates that the KS-reconstruction is the smoothest one, and thus shows the least amount of noise; one can see as well that the SS- and S-reconstruction have roughly the same quality, and that the B-reconstruction is by far the noisiest. For the smoothing length used, the KS-inversion does not resolve the maxima as well as the three finite-field kernel inversions do. The KS-reconstruction shows the well known artefacts at the boundary, i.e., local maxima at the corners of the field as well as local minima at the middle of the edges. The other reconstruction methods also show enhanced errors at the boundary; however, they are not systematic, but caused by increasing noise, since for positions near the CCD edge, the quantity $g$ is determined by averaging over fewer galaxies (in the extreme case, at a corner, one averages only over one fourth of the galaxy number that one averages for a point away from the boundary). In the reconstructions shown in Fig. 4, one can also see the impact of intrinsically very elliptical galaxies, see, e.g., the right front corner of the 'CCD'. Near that corner there is an intrinsically very elliptical source yielding a large estimate for the distortion; therefore, all reconstructions show a local maximum at this corner. For the KS-reconstruction, this local maximum is not so pronounced, because the reconstructed field is already strongly biased at the corners, and it appears that the KS-kernel is less sensitive to individual intrinsically elliptic sources than the three finite-field kernels.

**Fig. 4.** Surface and contour plots of $-K^{\mathrm{KS}}$, $-K^{\mathrm{SS}}$, $-K^{\mathrm{S}}$ and $-K^{\mathrm{B}}$ obtained by the inversion of the lens B using the parameters of reconstruction B2, i.e. $L = 7.5$, $N = 50$, $n_0 = 50$, $\rho = 0.2$, $\theta_0 = 0.34$ and $\theta_{\mathrm{sm}} = 0.11$; the contours differ by 0.1

On second sight, however, the impression that the KS-reconstruction is less noisy than the SS- and S-reconstruction must be modified. Namely, this impression is mainly due to the reconstruction near the boundary. As we stated before, due to the fact that we have less information about the distortion near the boundary, compared to points closer to the center of the field, we expect that *all* finite field inversions will be noisier close to the boundary. For the KS-reconstruction, the reduced information about the distortion near the boundary is substituted by artificially setting the shear to zero outside $\mathcal{U}$. Hence, one uses more (but wrong) information in the KS-inversion, which yields a smoother appearance of the reconstructed density field.

We next study the reconstruction with the four techniques for an ellipticity distribution with a width of $\rho = 0.3$, where more than 50 percent of the sources have an axis ratio smaller than 0.6. We have reconstructed the lens B for a galaxy density of $n_0 = 50$, ellipticity width $\rho = 0.3$ and outer smoothing length of $\theta_{\rm sm} = 0.11$, and we varied the inner smoothing length: $\theta_0 = 0.25$ (reconstruction B3), $\theta_0 = 0.3$ (reconstruction B4), $\theta_0 = 0.33$ (reconstruction B5), $\theta_0 = 0.51$ (reconstruction B6). The resulting power spectra can be seen in Fig. 5.

As expected, the power spectrum becomes larger if one compares a reconstruction for an ellipticity distribution with width $\rho = 0.2$ with that of a width of $\rho = 0.3$ using a fixed smoothing length, see, e.g., Fig. 3b and Fig. 5c: To first order, the power spectrum is roughly doubled for all inversion kernels; however, comparing Fig. 5c and Fig. 3b in more detail shows that the sensitivity to noise, introduced by the more elliptical source distribution, increases from the KS- to the SS- and S-reconstruction, and it is strongest in the B-reconstruction. Fig. 5 also shows how the power spectrum changes with increasing smoothing length, with all other parameters kept fixed. Generally, one needs a larger smoothing length for the SS-, S- and B-inversion than for the KS-inversion to lower the noise, where always the SS-kernel yields the best reconstruction of the three finite-field kernels. The short wave vector power spectrum for KS is increasing with larger smoothing length, since for KS this smoothing length already lowers the two maxima of the reconstructed $K$-field considerably. For the finite-field kernels this suppression of density peaks becomes important at larger smoothing length than for KS, presumably because for the KS-inversion one has to use the quantity $g$ itself, whereas for the finite-field kernels one needs the derivatives of $g$. Finally, using a very large smoothing length like $\theta_0 = 0.5$ or even larger yields a convergence of the power spectra of the 4 inversion types for large wave vectors: then, the reconstructions become all rather smooth, but small scale structures can no longer be reconstructed successfully. In Fig. 6 we demonstrate how the quality of reconstruction changes with decreasing and increasing galaxy density.

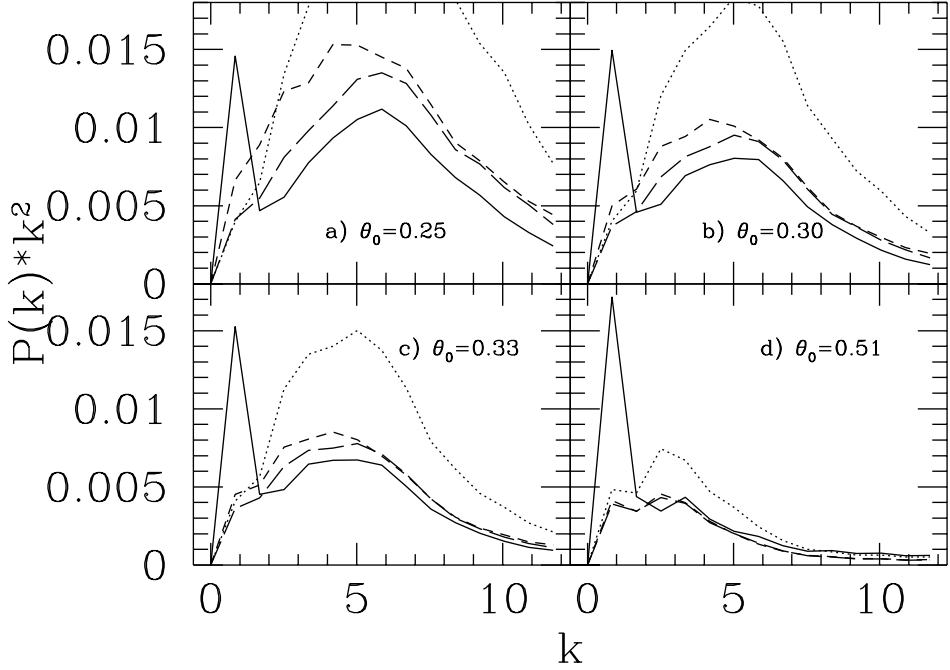

**Fig. 5.** Power spectra of the error field for reconstructing the lens B with $n_0 = 50$, $\rho = 0.3$, $\theta_{\rm sm} = 0.11$; the 'inner smoothing' length was: (a) $\theta_0 = 0.25$, (b) $\theta_0 = 0.3$, (c) $\theta_0 = 0.33$ and (d) $\theta_0 = 0.51$. In all 4 panels, the solid (long-dashed, short-dashed and dotted) line corresponds to the KS-power spectrum (SS-, S- and B-power spectrum). As expected, the smoothing length has to be increased compared to the case of $\rho = 0.2$ to get a noise-poor reconstruction, i.e. a small power spectrum.

Reducing the galaxy density from $n_0 = 50$ to $n_0 = 40$ means that the mean separation increases by about ten percent; hence, keeping the smoothing length constant (i.e., averaging over fewer galaxies to estimate $g$) yields an increase in the power spectrum for all inversion methods, as can be verified by comparison of Fig. 5b with Fig. 6a. However, the loss of reconstruction quality increases from the KS-reconstruction to the SS-, S- and B-reconstruction. An increase of the smoothing length to 0.4 and 0.5 can reduce the power spectrum again. Clearly, increasing the galaxy density reduces the power spectra, and the reconstruction with the KS-, SS- and S- technique become very simular, where, of course, the latter two do not show the spike of the power spectrum at small $k$; thus, at galaxy densities of 80 the SS- and S-reconstruction techniques are as good or superior to the KS-reconstruction on all scales.

If the ellipticity distribution of the sources is more extreme, $\rho = 0.4$, then for a galaxy density of 50, the three finite-field kernels yield only very noisy reconstructions, compare Fig. 5a and 7a. Using a large smoothing length of $\theta_{\rm sm} = 0.5$, see Fig. 7b, improves the reconstruction considerably and yields similar results as for the KS-reconstruction, but at the cost of small-scale resolution. The first maximum in the power spectra of the three finite-field kernels in Fig. 7b comes from the fact that with very elliptical sources, the

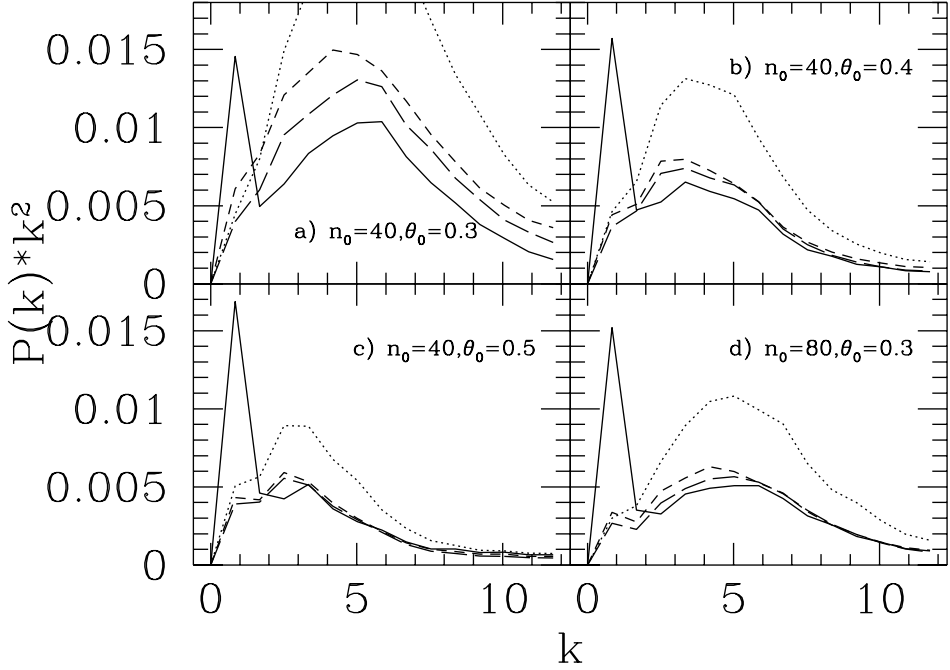

**Fig. 6.** Power spectra for the error field where the lens B was reconstructed, using a width of $\rho = 0.3$ in ellipticity distribution and $\theta_{\rm sm} = 0.11$; galaxy density and inner smoothing lenght were chosen as follows: (a) $n_0 = 40$, $\theta_0 = 0.3$, (b) $n_0 = 40$, $\theta_0 = 0.4$ (c) $n_0 = 40$, $\theta_0 = 0.5$ and (d) $n_0 = 80$, $\theta_0 = 0.3$. Reducing the galaxy density and not increasing the inner smoothing length yields a increase of the power spectrum for all 4 inversion techniques, compare Fig. 5b and Fig. 6a; the loss in quality of the reconstruction increases from the KS-, to the SS-, S- and B-reconstruction

reconstruction errors near the edges and especially at the corners become very large. In contrast to the KS-inversion, where the reconstruction at the outer parts of the CCD is systematically affected, one obtains for the finite-field kernels a 'wavy density field' at the outer parts. Fig. 7c demonstrates how much an increase in the galaxy density helps in the improvement of reconstruction quality if the galaxies are intrinsically very elliptical: compare 7a and 7c, where the parameters for which the power spectra are calculated differ only in the galaxy density. Also, as expected, if the galaxy density is large, e.g., $n_0 = 80$, a change in the ellipticity distribution towards more elliptical sources does not affect the reconstruction as much as for the case of a lower galaxy density (see the change in the power spectra from Fig. 6d to 7c, where for $n_0 = 80$, both $\rho$ and $\theta_0$ are increased from 0.3 to 0.4, and compare this to the change of the power spectra from Fig. 5b to 7a, where the galaxy density is 50, and also both $\rho$ and $\theta_0$ are increased from 0.3 to 0.4). We want to point out that the loss in reconstruction quality for the finite-field kernels for $\rho = 0.4$ compared to $\rho = 0.3$ is mainly caused by the increased fraction of galaxies which have an intrinsic axis ratio of 0.2 or smaller (see Fig. 2). However, since spiral galaxies have a finite disc, even an edge-on galaxy will have an axis ratio of 0.2 or larger; therefore, an ellipticity distribution with a width of $\rho = 0.4$ appears quite unrealistic. Despite of this, Fig. 7 shows that cluster inversion can be done successfully even in the presence of very

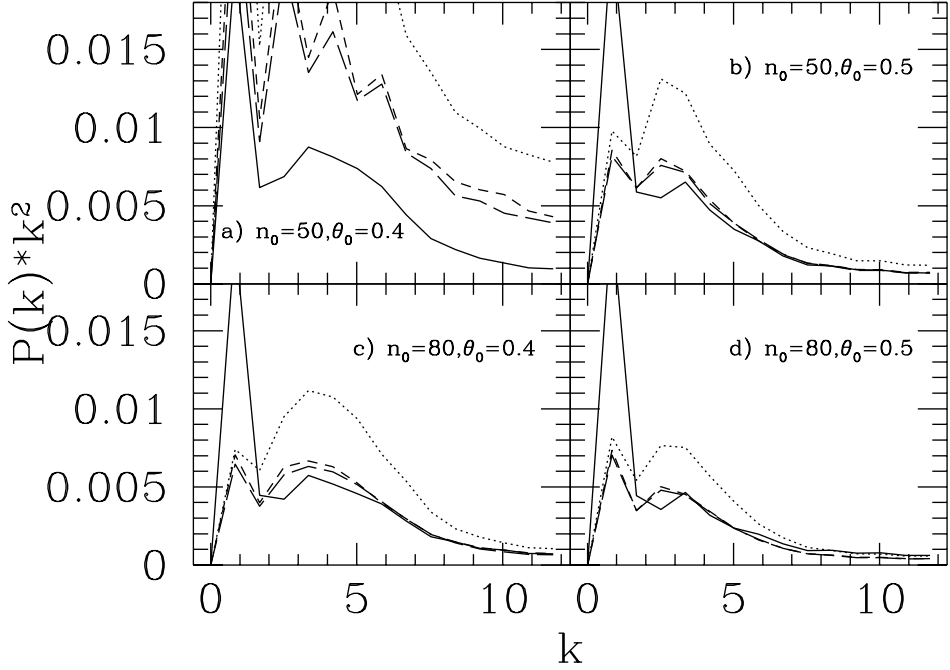

**Fig. 7.** Power spectra for the extreme case of an ellipticity distribution with a width of $\rho = 0.4$; this ellipticity distribution is not realistic since quite many galaxies with an axis ratio of 0.2 or smaller are generated; however, the power spectra demonstrate that the KS-inversion method can extract the distortion quite well even in the presence of extremely elliptical galaxies. The galaxy density and inner smoothing length have been varied as follows: (a) $n_0 = 50$, $\theta_0 = 0.4$, (b) $n_0 = 50$, $\theta_0 = 0.5$, (c) $n_0 = 80$, $\theta_0 = 0.4$ and (d) $n_0 = 80$, $\theta_0 = 0.5$; in all four panels, the outer smoothing length was fixed as $\theta_{\rm sm} = 0.11$. As in the preceding figures, solid (long-dashed, short-dashed and dotted) lines denote KS- (SS-, S- and B-) power spectra

elliptical sources, if only the observed galaxy density is sufficiently high.

To see what happens if the inverted mass distribution is not centered on the center of the CCD, we reconstruct the mass distribution of lens A. This does not mean that we expect an observer to position the telescope on purpose in such a way; however, it demonstrates how large the systematic error of the KS-reconstruction can become in unfavourable cases; in addition, we note that in some cases (see Bonnet, Mellier & Fort 1994) where bright stars are in the foreground, one sometimes must position the CCD in a non-optimal position for cluster inversion.

Fig. 8 demonstrates that the KS-reconstruction quality decreases very strongly, and that for the finite-field kernels this reduces the reconstruction quality only slightly (due to our smoothing procedure, the noise of the reconstruction at the boundary is larger). A more realistic observational situation is simulated in the reconstruction of lens C and D. It may happen that someone tries to invert a mass profile of two groups of galaxies (lens D); then, a nearby additional dark mass distribution (lens C) will lead to large errors in the KS-reconstruction. (Such dark mass distributions do exist as has been proven by the detection of shear fields around bright 1-Jansky quasars, caused by 'dark' mass

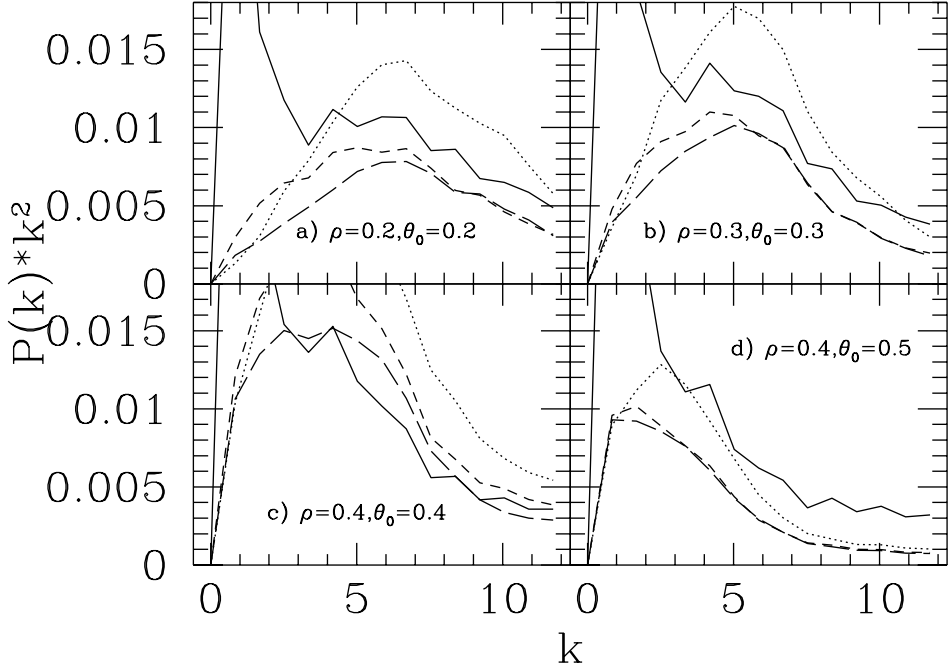

**Fig. 8.** Power spectra for the error field for the reconstruction of the lens A: in all panels, $n_0 = 50$ and $\theta_{\rm sm} = 0.11$; the ellipticity width and inner smoothing length are varied. Since the mass profiles of lens models B and A are quite similar, these power spectra are shown to demonstrate the increase of reconstruction error for the case where the center of the mass distribution is shifted to an edge of the CCD field. Hence, compare Fig. 8a with 3a, 8b with 5b, 8c with 7a and 8d with 7b

distributions; B. Fort, private communication.)

For lens D, the KS-reconstruction yields a good estimate, since the shear outside the data field, which is set to zero in the KS-inversion, then is indeed very small. Hence, the spike at small wave vectors in Fig. 9a is very small, and the KS-, SS- and S-estimates of the mass distribution are then almost equally good (Fig. 9a). The additional mass distribution at the boundary of the CCD in lens C, however, yields again a prominent rise of the spike (Fig. 9b). Hence, also for the reconstruction of a very weak cluster or groups of galaxies, it is profitable to use a finite-field kernel inversion.

In order to demonstrate at which positions the errors in the 4 different inversion methods are particulary large, we have plotted the root mean square of the reconstruction error as a function of position, see Fig. 10.

The rms error was obtained by 500 simulations of reconstruction of lens B, using a galaxy density of 50, a width in the ellipticity distribution of $\rho = 0.3$, an inner and outer smoothing length of 0.3 and 0.11. The corresponding power spectra of the error fields are those in Fig. 5b. The rms error at a position $\boldsymbol{\theta}$ is defined as follows

$$\Delta K(t) := \sqrt{<(K_{\rm shift}(\boldsymbol{\theta}) - K_{\rm true}(\boldsymbol{\theta}))^2>} \quad ,$$

where $K_{\rm shift}$ is the reconstructed $\tilde{K}$-field obtained for the estimator considered, shifted

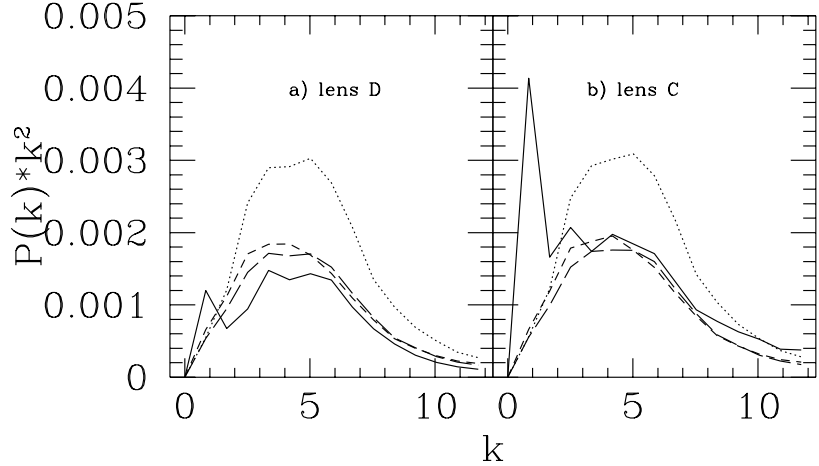

**Fig. 9.** Power spectra of the error field for reconstructed lens D (a) and lens C (b); as in previous plots, the solid line (long-dashed, short-dashed and dotted lines) belong to the KS- (SS-, S- and B-) power spectrum. (a) For a weak mass distribution at the center of the CCD field, the KS-reconstruction yields a good estimate, where the finite-field kernel reconstructions are a little bit noisier on small scales; (b) if there is an additional mass at the boundary of the CCD, the quality of reconstruction decrases considerably for the KS-reconstruction, whereas it stays approximately the same for the finite-field kernel reconstructions

such that the sum of $K_{\rm shift}$ over the data-field equals that of $K_{\rm true}$. The rms error field for the KS-reconstruction is shown in the upper left panel of Fig. 10: i) the reconstruction error is largest at the corners where it exeeds 0.2; ii) the rms-error field shows a maximum at the center of the data field; iii) 'behind' the central maximum, there is a strong increase of the error towards the edge of the data-field. The reconstruction error has to be large at the corners and it is large along those edges where the shear field is still high (this explains why the rms error is not as large on the remaining three edges). However, we want to point out that the comparison done in Fig. 10 is not 'fair', due to the normalization of the $K$-field. Since the normalization is affected by the boundary artefacts, it may increase the error in the middle of the data field. However, we have not come up with a 'fairer' normalization. For this reason, the comparison of Fig. 10a with the other panels should be considered with care.

The finite-field kernels rms error field are qualitatively similar to each other: at the edges and corners the rms error is also increased, but contrary to the KS-reconstruction, the rms error arises only from the increased noise (smoothing over fewer galaxies) in the mass distribution at these positions. This can partly be avoided by a more elaborated smoothing procedure, which, however, will not be discussed in this paper.

A comparison of the rms errors of the finite-field kernels with each other yields the same ranking for these inversion techniques as from considering the power spectra: the size of the rms error increases from SS-, to the S- and B-reconstruction. In particular, the error of the SS-inversion seems to be spatially flatter than for the other two finite-field

**Fig. 10.** Root mean square error of the reconstruction as a function of position for the KS-, SS-, S- and B-reconstruction. The number of simulations is 500, $n_0 = 50$, $\rho = 0.3$, $\theta_0 = 0.3$ and $\theta_{\rm sm} = 0.11$; the corresponding power spectra for the error field are displayed in Fig. 5b. The vertical axis is drawn from 0 to .2; the contour lines are in the same interval, with a step size of 0.0125; the size of the CCD field is 7.5 arcminutes, as usual

inversion techniques.

# 6 Discussion

In this paper we have derived a cluster inversion formula which yields the surface mass density distribution from observations of the image distortion on a finite data (CCD) field. This new inversion equation is obtained by convolving the vector field $\mathbf{u}(\boldsymbol{\theta})$, which is obtained from the observable $g(\boldsymbol{\theta})$ and its derivatives – see Eq. (3.2) – with a kernel $\mathbf{H}(\boldsymbol{\theta}',\boldsymbol{\theta})$, which in turn is obtained by solving a Laplace-like equation with Neumann boundary conditions. We have explicitly constructed the kernel $\mathbf{H}$ for a rectangular field, as described in the appendix (for a circular field, the kernel $\mathbf{H}$ is known in closed form; for more general field geometries, $\mathbf{H}$ has to be calculated numerically). We then have compared this new formula with the nonlinear version of the KS inversion equation and two other finite-field inversion formulae derived in Paper III, and KSFWB and Bartelmann (1995), respectively, by applying these various methods to synthetic data and performing a spectral analysis of the deviations of the resulting mass profile from the original mass distribution. The main finding of this comparison is that our new formula in all cases considered is better than the two other finite-field kernels, and at worst only slightly more noisy than the KS reconstruction; however, this slight increase in noise is more than compensated by the removal of boundary artefacts with which the KS method is burdened. If the mass distribution extends near to the field boundaries, then our new method yields reconstructions which deviate less, even on small scales, from the input mass distribution than that obtained from the KS formula, since then the latter can be seriously hampered by boundary effects. This effect becomes even more pronounced if rectangular data fields with side ratios not close to one are considered (which we have not done in this paper, but which was demonstrated in Paper III). In most cases, our new reconstruction kernel yields only slightly better results than the method of Paper III, whereas in nearly all cases it is considerably better than that of KSFWB and Bartelmann (1995). We also want to point out that the noise of our new reconstruction method seems to be uniform over the data field, except that it is slightly larger at the boundary (which must be the case since there is less information about the distortion near the boundary), whereas the other three inversions have their noise more concentrated towards the boundary.

The finite-field inversions S and B were derived from line integrations of the gradient of the surface mass density (3.1) or (3.2) and averaging over many different curves l; hence, they are special cases of the more general inversion formula (3.7). This equation leaves a lot of freedom (or rather: arbitrariness), i.e., the choice of the weight function $w(\boldsymbol{\theta}_0)$ and the choice of the curves l connecting $\boldsymbol{\theta}_0$ and $\boldsymbol{\theta}$. In Paper III, these integration contours were chosen such that the resulting inversion equations agrees locally with that of the KS formula; in KSFWB and Bartelmann (1995), this is not the case. As argued in Paper III, this requirement appears reaonable due to the singularity of the kernel for $\boldsymbol{\theta}'$ close to $\boldsymbol{\theta}$, and we attribute the considerably worse noise properties of the B inversion relative to the S inversion to this fact. We also note that we have used the B inversion in the form (3.8), not in its integrated form; some test calculations have shown that the latter one yields results which are noisier than those obtained from (3.8), due to the unsmooth behaviour of $R(\boldsymbol{\theta}', \boldsymbol{\theta})$ – see Eq. (3.9) – in the direction towards the corners. It may well be that for field geometries with smooth boundary curves this difference will vanish; however, we think that the local anisotropy of the kernel $\mathbf{H}^{\mathrm{B}}$ for $\boldsymbol{\theta}' \to \boldsymbol{\theta}$ is the main cause for its increased noise relative to the S inversion. The choice of the integration curves for the S inversion in Paper III appears artifical at first sight, but has proven to be reasonable. However, for both the S and the B inversion (in their present form) it is essential that the data field is convex.

The rationall behind our new inversion method was the consideration that the 'observed' vector field $\mathbf{u}(\boldsymbol{\theta})$ will in general not be rotation-free. Our new kernel was derived by requiring that any rotational component of $\mathbf{u}$ is filtered out, and that the decomposition of the field $\mathbf{u}$ into a gradient and a rotational part is such that the latter is minimized in a particular sense (see Sect. 4). These two requirements then specify the integral kernel $\mathbf{H}(\boldsymbol{\theta}', \boldsymbol{\theta})$ uniquely. In particular, the generalization to a different field geometry is reduced to the solution of a Laplace equation with Neumann boundary conditions, and non-convex data fields can be treated with the same method. We shall treat other geometries elsewhere, in particular that of the (non-convex) WFPC-2 field on board HST.

The fact that our new inversion method has not significantly worse noise properties than KS is surprising at first sight, given the fact that it needs the data in a differentiated form. Of course, the field $\mathbf{u}$ is more noisy that the input field $g$; however, the integration (4.3) appears to cure this additional noise caused by finite differencing.

We have not tried to find optimal smoothing procedures in this paper, and in fact used the same smoothing procedures for all four inversion techniques. It is likely that each of these inversions has its own optimal smoothing scale, but that will depend strongly on the quality of the data available. Hence, the choice of the smoothing procedure has to be made after the actual data set is obtained.

Looking at the power spectra in Sect. 5 one might suggest alternative inversion methods by combining our new formula with that of KS. This could be done either by a Fourier decomposition of the KS reconstruction and the SS reconstruction, and then using the SS components for the large scales and the KS components for small scales. Alternatively, one could use the SS reconstructed density field to calculate its shear outside the data field and then apply the KS inversion, using the observed field $g$ inside the data field and the calculated field $g$ outside the data field. Several more variants of this kind can be thought of. On the other hand, it remains to be seen how useful the method of Broadhurst, Taylor & Peacock (1995) and, in particular, that of Bartelmann & Narayan (1995) will be. Certainly, both of these new methods (which make use of the magnification effect which is completely neglected in the inversion techniques discussed here) will yield additional information about the surface mass density of the cluster; in particular, the additive constant in the $\kappa$-distribution can probably be determined from these methods. It will be difficult to incorporate this additional information into inversion formulae of the type discussed here, and one will have to think about more general inversion techniques which can make use of all available information. Most likely, this will lead to some kind of maximum-likelihood approach for a parametrized mass distribution. Such a method should also take into account the redshift distribution of the sources, and in fact should be able to determine that distribution if several clusters at different redshifts are simultaneously fitted. A very first step in this direction was taken in Smail, Ellis & Fitchett (1995).

It is a pleasure to acknowledge very useful discussions with C. Seitz, N. Kaiser, G. Squires, H.-W. Rix, and, in particular, S. White, who suggested a geometrical method to solve our Neumann problem, as done in the appendix, and M. Bartelmann and J. Ehlers for their most detailed comments on the manuscript. This work was supported in part by the Sonderforschungsbereich SFB 375-95 of the Deutsche Forschungsgemeinschaft.

# Appendix

Here we want to derive the solution for the 'Poisson' equation

$$\Delta \mathcal{L}(\boldsymbol{\theta}', \boldsymbol{\theta}) = -\delta(\boldsymbol{\theta} - \boldsymbol{\theta}') + \frac{1}{A} \quad , \tag{A1}$$

for $\boldsymbol{\theta}$ inside the region $\mathcal{U}$ with area $A$, together with the boundary condition

$$\mathbf{H}(\boldsymbol{\theta}', \boldsymbol{\theta}) \cdot \mathbf{n}(\boldsymbol{\theta}') = \frac{\mathrm{d}}{\mathrm{d}\mathbf{n}'} \mathcal{L}(\boldsymbol{\theta}', \boldsymbol{\theta}) = 0 \tag{A2}$$

for all points $\boldsymbol{\theta}'$ on the boundary $\partial \mathcal{U}$ of $\mathcal{U}$. Here,

$$\mathbf{H}(\boldsymbol{\theta}', \boldsymbol{\theta}) \equiv \nabla \mathcal{L}(\boldsymbol{\theta}', \boldsymbol{\theta}) \quad , \tag{A3}$$

all differential operators in this appendix are to be taken with respect to $\boldsymbol{\theta}'$, and $\mathbf{n}$ is the outward-directed normal vector on the boundary of $\mathcal{U}$. The existence of a solution of this Neumann problem is guaranteed by the compatibility condition

$$\int \mathrm{d}^2 \boldsymbol{\theta}' \; \Delta \mathcal{L}(\boldsymbol{\theta}', \boldsymbol{\theta}) = 0 \quad , \tag{A4}$$

(see, e.g., Garabedian 1986, p. 230), and it is unique up to an additive constant, which implies that the solution for $\mathbf{H}$ is unique. Here we shall derive the solution for two special geometries, a circle of radius $R$ and a rectangle (which is the most relevant case due to the shape of CCDs and most, though not all, CCD arrays).

For a circle of radius R, the solution of the Neumann problem can be found in textbooks (e.g., Garabedian 1986, p. 248); we only quote the result for $\mathbf{H}$:

$$\mathbf{H}(\boldsymbol{\theta}', \boldsymbol{\theta}) = \frac{1}{2\pi} \left( \frac{\boldsymbol{\theta}'}{R^2} + \frac{\boldsymbol{\theta} - \boldsymbol{\theta}'}{|\boldsymbol{\theta} - \boldsymbol{\theta}'|^2} + \frac{R^2 \boldsymbol{\theta}/|\boldsymbol{\theta}|^2 - \boldsymbol{\theta}'}{\left| R^2 \boldsymbol{\theta}/|\boldsymbol{\theta}|^2 - \boldsymbol{\theta}' \right|^2} \right) \quad . \tag{A5}$$

The validity of this solution can be easily checked.

Consider next a rectangle with sidelengths $L_1$ and $L_2$. In this case, the solution of our boundary value problem can be found from a geometrical consideration. First of all, we note that the function

$$\mathcal{G}(\boldsymbol{\theta}', \boldsymbol{\theta}) := \frac{\ln|\boldsymbol{\theta}' - \boldsymbol{\theta}|}{2\pi} \tag{A6}$$

satisfies $\Delta \mathcal{G}(\boldsymbol{\theta}', \boldsymbol{\theta}) = \delta(\boldsymbol{\theta} - \boldsymbol{\theta}')$. Hence, the vector field

$$\mathbf{h}(\boldsymbol{\theta}', \boldsymbol{\theta}) = -\nabla \mathcal{G}(\boldsymbol{\theta}', \boldsymbol{\theta}) + \int_\mathcal{U} d^2\vartheta \, \frac{\nabla \mathcal{G}(\boldsymbol{\vartheta}, \boldsymbol{\theta}')}{A} = \frac{1}{2\pi} \frac{\boldsymbol{\theta} - \boldsymbol{\theta}'}{|\boldsymbol{\theta} - \boldsymbol{\theta}'|^2} + \int_\mathcal{U} d^2\vartheta \, \frac{1}{2\pi A} \frac{\boldsymbol{\theta}' - \boldsymbol{\vartheta}}{|\boldsymbol{\theta}' - \boldsymbol{\vartheta}|^2} \quad (A7)$$

satisfies the 'Poisson' equation (A1), i.e., $\nabla \cdot \mathbf{h}(\boldsymbol{\theta}', \boldsymbol{\theta}) = -\delta(\boldsymbol{\theta} - \boldsymbol{\theta}') + 1/A$, with $A = L_1 L_2$. The vector field $\mathbf{h}(\boldsymbol{\theta}', \boldsymbol{\theta})$ has the geometric interpretation of a deflection angle at point $\boldsymbol{\theta}'$ caused by a homogeneous mass distribution of surface mass density $1/(2A)$ inside the rectangle, and zero surface mass density outside, plus that of a point 'mass' at position $\boldsymbol{\theta}$ in $\mathcal{U}$ with negative mass $-1/2$. Of course, $\mathbf{h}$ does not satisfy the boundary condition (A2). Let us define the points $\boldsymbol{\theta}^{(ij)}$ for $i \neq 0$, $j \neq 0$ by $\boldsymbol{\theta}^{(ij)} = \left(\theta_1^{(i)}, \theta_2^{(j)}\right)$, with ($k = 1, 2$)

$$\theta_k^{(i)} = \begin{cases} \theta_k + (i-1)L_k & \text{for } i \text{ odd}, \\ iL_k - \theta_k & \text{for } i \text{ even}, \end{cases} \quad (A8)$$

for $i \geq 1$, and with $\theta_k^{(i)} = -\theta_k^{(-i)}$ for negative $i$. Furthermore, we define the rectangular regions $\mathcal{U}^{(ij)}$ to have sidelengths $L_1$, $L_2$, and to be centered on $\mathbf{c}^{(ij)} = \left(c_1^{(i)}, c_2^{(j)}\right)$, with

$$c_k^{(i)} = (2i-1)L_k/2 \quad \text{for} \quad i \geq 1 \quad,$$
$$c_k^{(i)} = -c_k^{(-i)} \quad \text{for} \quad i \leq -1 \quad.$$

Then, $\boldsymbol{\theta}^{(11)} = \boldsymbol{\theta}$, $\mathcal{U}^{(11)} = \mathcal{U}$, $\boldsymbol{\theta}^{(-i,-j)} = -\boldsymbol{\theta}^{(ij)}$, etc. Note that the point $\boldsymbol{\theta}^{(ij)}$ lies in $\mathcal{U}^{(ij)}$. Fig.11 illustrates the geometrical arrangement of the points $\boldsymbol{\theta}^{(ij)}$ and the rectangles $\mathcal{U}^{(ij)}$. In analogy to (A7), we define the vector fields, for $i \neq 0$, $j \neq 0$,

$$\mathbf{h}^{(ij)}(\boldsymbol{\theta}', \boldsymbol{\theta}) = -\nabla \mathcal{G}\left(\boldsymbol{\theta}', \boldsymbol{\theta}^{(ij)}\right) + \int_{\mathcal{U}^{(ij)}} d^2\vartheta \, \frac{\nabla \mathcal{G}(\boldsymbol{\vartheta}, \boldsymbol{\theta}')}{A} \quad, \quad (A9)$$

which can be interpreted as the deflection angle at the point $\boldsymbol{\theta}'$ caused by a uniform surface mass density of magnitude $1/(2A)$ in the rectangle $U^{(ij)}$ plus the deflection by a negative point mass $-1/2$ at $\boldsymbol{\theta}^{(ij)}$. We then have, for $\boldsymbol{\theta} \in \mathcal{U}$, $\boldsymbol{\theta}' \in \mathcal{U}$, that $\mathbf{h}^{(11)}(\boldsymbol{\theta}', \boldsymbol{\theta}) = \mathbf{h}(\boldsymbol{\theta}', \boldsymbol{\theta})$, and $\nabla \cdot \mathbf{h}^{(ij)}(\boldsymbol{\theta}', \boldsymbol{\theta}) = 0$ for $(i, j) \neq (1, 1)$. Since the geometrical arrangement of rectangles and points is symmetric with respect to a reflection around the lines $\theta_1' = mL_1$, and $\theta_2' = nL_2$, the sum

$$\mathbf{H}(\boldsymbol{\theta}', \boldsymbol{\theta}) = \sum_{|i|=1}^{N_x} \sum_{|j|=1}^{N_y} \mathbf{h}^{(ij)}(\boldsymbol{\theta}', \boldsymbol{\theta}) \quad (A10)$$

satisfies the Poisson equation $\nabla \cdot \mathbf{H}(\boldsymbol{\theta}', \boldsymbol{\theta}) = -\delta(\boldsymbol{\theta} - \boldsymbol{\theta}') + 1/A$ for $\boldsymbol{\theta}, \boldsymbol{\theta}' \in \mathcal{U}$, and also satisfies the boundary condition (A2), due to symmetry, provided the sum in (A10) is extended to infinity in all directions. However, special care has to be taken by performing this sum: the magnitude of $\mathbf{h}^{(ij)}$ decreases as $1/r^2$, where $r$ is the distance of the rectangle $\mathcal{U}^{(ij)}$ from the origin (and not as $1/r$, since the 'monopole' contribution of the negative point mass is canceled by the uniform surface mass density, i.e., the total mass inside $\mathcal{U}^{(ij)}$ vanishes),

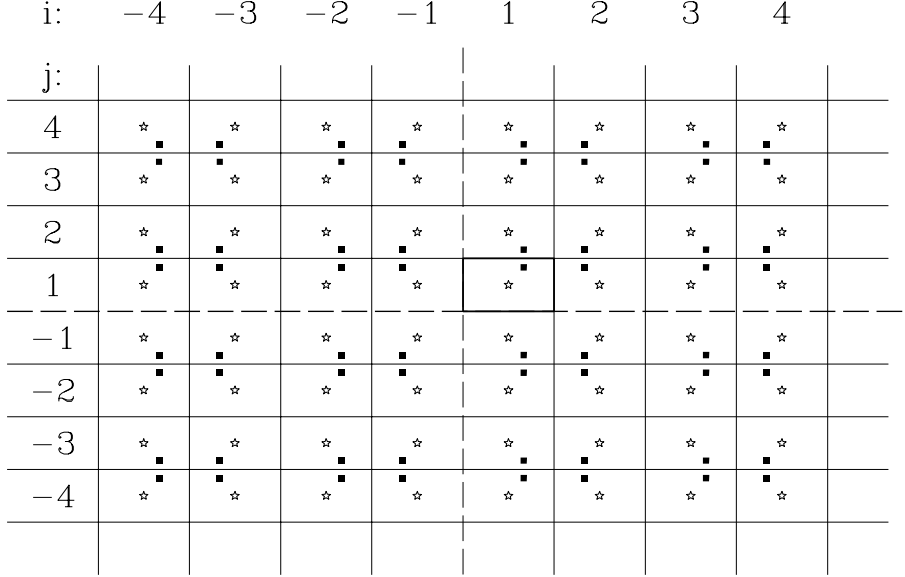

**Fig. 11.** The arrangement of the fields $\mathcal{U}^{(ij)}$ and angles $\boldsymbol{\theta}^{(ij)}$ is scetched for $-4 \leq i, j \leq 4$, $i, j \neq 0$. The stars denote the positions $\mathbf{c}^{(ij)} = \left(c_1^{(i)}, c_2^{(j)}\right)$ of the centers of a fields $\mathcal{U}^{(ij)}$, the filled squares denote the points $\boldsymbol{\theta}^{(ij)}$. The boundary of the field $\mathcal{U}^{(11)} = \mathcal{U}$ is marked with a thick solid line, the dashed lines are the coordinate axes

and the two-dimensional sum can diverge for $N_x, N_y \to \infty$ if this limiting process is not defined carefully.

In order to see how this limiting process should be done (in particular for the numerical realization), consider the contribution of the four rectangles around the origin,

$$\mathbf{w}(\boldsymbol{\theta}', \boldsymbol{\theta}) := \sum_{i,j=\pm 1} \mathbf{h}^{(ij)}(\boldsymbol{\theta}', \boldsymbol{\theta}) = -\sum_{i,j=\pm 1} \nabla \mathcal{G}(\boldsymbol{\theta}, \boldsymbol{\theta}^{(ij)}) + \mathbf{F}(\boldsymbol{\theta}', L_1, L_2) \quad , \tag{A11}$$

where

$$\mathbf{F}(\mathbf{x}; A, B) = \int_{-A}^{A} \mathrm{d}x_1' \int_{-B}^{B} \mathrm{d}x_2' \frac{\mathbf{x} - \mathbf{x}'}{|\mathbf{x} - \mathbf{x}'|^2} \quad . \tag{A12a}$$

Evaluating the integrals, this becomes

$$F_1(\mathbf{x}; A, B) = \frac{B + x_2}{2} \ln \frac{(B + x_2)^2 + (A + x_1)^2}{(B + x_2)^2 + (A - x_1)^2} + \frac{B - x_2}{2} \ln \frac{(B - x_2)^2 + (A + x_1)^2}{(B - x_2)^2 + (A - x_1)^2}$$

$$+ (A + x_1) \left[ \arctan\left(\frac{B + x_2}{A + x_1}\right) + \arctan\left(\frac{B - x_2}{A + x_1}\right) \right]$$

$$- (A - x_1) \left[ \arctan\left(\frac{B + x_2}{A - x_1}\right) + \arctan\left(\frac{B - x_2}{A - x_1}\right) \right] ,$$

$$F_2(\mathbf{x}; A, B) = \frac{A + x_1}{2} \ln \frac{(A + x_1)^2 + (B + x_2)^2}{(A + x_1)^2 + (B - x_2)^2} + \frac{A - x_1}{2} \ln \frac{(A - x_1)^2 + (B + x_2)^2}{(A - x_1)^2 + (B - x_2)^2}$$

$$+ (B + x_2) \left[ \arctan\left(\frac{A + x_1}{B + x_2}\right) + \arctan\left(\frac{A - x_1}{B + x_2}\right) \right]$$

$$- (B - x_2) \left[ \arctan\left(\frac{A + x_1}{B - x_2}\right) + \arctan\left(\frac{A - x_1}{B - x_2}\right) \right] .$$

$$(A12b)$$

We note the following expansion for $\mathbf{F}$:

$$F_1(\mathbf{x}, A, B) = 4 \arctan\left(\frac{B}{A}\right) x_1 + \frac{4AB}{3(A^2 + B^2)^2} x_1 \left(x_1^2 - 3x_2^2\right)$$

$$+ \frac{4AB(A^2 - B^2)}{5(A^2 + B^2)^4} x_1 \left(x_1^4 - 10x_1^2 x_2^2 + 5x_2^4\right) + \mathcal{O}\left(\frac{|\mathbf{x}|^7}{A^6}\right) ,$$

$$F_2(\mathbf{x}, A, B) = 4 \arctan\left(\frac{A}{B}\right) x_2 + \frac{4AB}{3(A^2 + B^2)^2} x_2 \left(x_2^2 - 3x_1^2\right)$$

$$- \frac{4AB(A^2 - B^2)}{5(A^2 + B^2)^4} x_2 \left(5x_1^4 - 10x_1^2 x_2^2 + x_2^4\right) + \mathcal{O}\left(\frac{|\mathbf{x}|^7}{A^6}\right) ,$$

$$(A12c)$$

$$F_1(\mathbf{x}, A, B) = \frac{4AB x_1}{|\mathbf{x}|^2} \left[ 1 + \frac{A^2 - B^2}{3} \frac{x_1^2 - 3x_2^2}{|\mathbf{x}|^4} \right.$$

$$+ \frac{3A^4 - 10A^2 B^2 + 3B^4}{15} \frac{x_1^4 - 10x_1^2 x_2^2 + 5x_2^4}{|\mathbf{x}|^8} + \mathcal{O}\left.\left(\frac{A^6}{|\mathbf{x}|^6}\right) \right] ,$$

$$F_2(\mathbf{x}, A, B) = \frac{4AB x_2}{|\mathbf{x}|^2} \left[ 1 + \frac{A^2 - B^2}{3} \frac{3x_1^2 - x_2^2}{|\mathbf{x}|^4} \right.$$

$$+ \frac{3A^4 - 10A^2 B^2 + 3B^4}{15} \frac{5x_1^4 - 10x_1^2 x_2^2 + x_2^4}{|\mathbf{x}|^8} + \mathcal{O}\left.\left(\frac{A^6}{|\mathbf{x}|^6}\right) \right] ,$$

The leading order term of $\mathbf{w}(\boldsymbol{\theta}', \boldsymbol{\theta})$ for $|\boldsymbol{\theta}'| \to \infty$ behaves like $|\boldsymbol{\theta}'|^{-3}$, because the dipole of the mass distribution inside these four rectangles vanishes. We will term this collection of four rectangles (and four negative point masses) a 'big rectangle'. Consider now this big rectangle to be centered on a point $\mathbf{C}^{(ij)} = (2iL_1, 2jL_2)$; then its contribution to the sum in (A10) will be $\mathbf{w}\left(\boldsymbol{\theta}' - \mathbf{C}^{(ij)}, \boldsymbol{\theta}\right)$. Even better behaved for $|i|, |j| \to \infty$ is the sum of four such big rectangles,

$$\mathbf{W}^{(i,j)}(\boldsymbol{\theta}',\boldsymbol{\theta}) = \sum_{\pm,\pm} \mathbf{w}\left(\boldsymbol{\theta}' - \mathbf{C}^{(\pm i, \pm j)}, \boldsymbol{\theta}\right) \quad , \tag{A13}$$

defined for $i,j \geq 1$, since the quadrupole of these four big rectangles also vanishes.

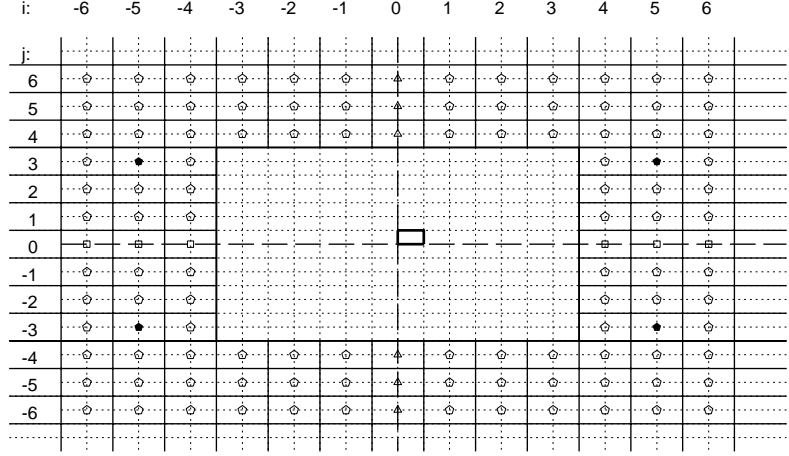

**Fig. 12.** This figure shows the arrangement of the 'big rectangles': the long-dashed lines mark the axis of the coordinate system, the thick solid line the boundary of the field $\mathcal{U} = U^{(1,1)}$ as in Fig. 11. The short-dashed lines incidate the 'small rectangles' $\mathcal{U}^{(i,j)}$; the big rectangles are drawn with solid lines, where their centers are denoted by triangles on the $x_2$-axis, squares on the $x_1$-axis, and pentaga elsewhere. Four of the pentaga are filled, to indicate how the sum over 4 'big rectangles', defined in equation (A13), has to be taken. The values for $i$ and $j$ at the left and the top of the figures denote the indices of the 'big rectangles' and their centers $C^{(ij)}$

An efficient way to perform the sum in (A10) is thus to split it into an 'inner' and an 'outer' part,

$$\mathbf{H}(\boldsymbol{\theta}',\boldsymbol{\theta}) = \mathbf{H}^{\text{in}}(\boldsymbol{\theta}',\boldsymbol{\theta}) + \mathbf{H}^{\text{out}}(\boldsymbol{\theta}',\boldsymbol{\theta}) \quad , \tag{A14}$$

where

$$\mathbf{H}^{\text{in}}(\boldsymbol{\theta}',\boldsymbol{\theta}) = \frac{1}{2\pi} \sum_{|i|=1}^{N_x} \sum_{|j|=1}^{N_y} \frac{\boldsymbol{\theta}^{(ij)} - \boldsymbol{\theta}'}{\left|\boldsymbol{\theta}^{(ij)} - \boldsymbol{\theta}'\right|^2} + \frac{1}{2\pi A}\mathbf{F}(\boldsymbol{\theta}'; N_x L_1, N_y L_2) \quad , \tag{A15}$$

and we have combined the contributions of the second term in (A9) from the inner part, and the function $\mathbf{F}$ was defined in (A12). We will choose $N_x$ and $N_y$ to be odd. Then, the outer part of $\mathbf{H}$ is evaluated as a sum over the $\mathbf{W}$'s,

$$\mathbf{H}^{\text{out}}(\boldsymbol{\theta}', \boldsymbol{\theta}) = \frac{1}{2} \left[ \sum_{i=(N_x+1)/2}^{M_x} \mathbf{W}^{(i0)}(\boldsymbol{\theta}', \boldsymbol{\theta}) + \sum_{j=(N_y+1)/2}^{M_y} \mathbf{W}^{(0j)}(\boldsymbol{\theta}', \boldsymbol{\theta}) \right]$$
$$+ \left[ \sum_{i=1}^{(N_x-1)/2} \sum_{j=(N_y+1)/2}^{M_y} \mathbf{W}^{(ij)}(\boldsymbol{\theta}', \boldsymbol{\theta}) + \sum_{i=(N_x+1)/2}^{M_x} \sum_{j=1}^{M_y} \mathbf{W}^{(ij)}(\boldsymbol{\theta}', \boldsymbol{\theta}) \right] . \quad (A16)$$

Using (A12c), we can now expand the functions $\mathbf{W}^{(ij)}$ in terms of $1/\left|\mathbf{C}^{(ij)}\right|$, and obtain

$$2\pi \mathbf{W}^{(ij)}(\boldsymbol{\theta}', \boldsymbol{\theta}) = (q_1 v_1 + q_2 v_2 + q_3 v_3) \begin{pmatrix} -\theta_1' \\ \theta_2' \end{pmatrix}$$
$$+ (10 q_2 v_1 + 7 q_3 v_2) \begin{pmatrix} -\theta_1' \left( \theta_1'^2 - 3\theta_2'^2 \right) \\ \theta_2' \left( 3\theta_1'^2 - \theta_2'^2 \right) \end{pmatrix} \qquad (A17)$$
$$+ 21 q_3 v_1 \begin{pmatrix} -\theta_1' \left( \theta_1'^4 - 10\theta_1'^2 \theta_2'^2 + 5\theta_2'^4 \right) \\ \theta_2' \left( 5\theta_1'^4 - 10\theta_1'^2 \theta_2'^2 + \theta_2'^4 \right) \end{pmatrix} + \cdots ,$$

with
$$\begin{aligned}
v_1 &= L_1^2 - L_2^2 - 3(\theta_1^2 - \theta_2^2) \quad, \\
v_2 &= 3(L_1^4 + L_2^4) - 10(L_1 L_2)^2 - 15\left(\theta_1^4 + \theta_2^4 - 6\theta_1^2 \theta_2^2\right) \quad, \\
v_3 &= 3\left[L_1^6 - L_2^6 - 7(L_1 L_2)^2 \left(L_1^2 - L_2^2\right)\right] - 21\left[\theta_1^6 - \theta_2^6 - 15(\theta_1 \theta_2)^2 \left(\theta_1^2 - \theta_2^2\right)\right] \quad, \\
q_1 &= \frac{16 \left(C_1^4 - 6 C_1^2 C_2^2 + C_2^4\right)}{|\mathbf{C}|^8} \quad, \\
q_2 &= \frac{16 \left[C_1^6 - 15(C_1 C_2)^2 \left(C_1^2 - C_2^2\right) - C_2^6\right]}{3 |\mathbf{C}|^{12}} \quad, \\
q_3 &= \frac{16 \left[C_1^8 + C_2^8 - 28(C_1 C_2)^2 \left(C_1^4 + C_2^4\right) + 70(C_1 C_2)^4\right]}{3 |\mathbf{C}|^{16}} \quad,
\end{aligned} \qquad (A18)$$

where we have written $\mathbf{C} \equiv \mathbf{C}^{(ij)}$. Since we are interested in the field $\mathbf{H}$ only for $\boldsymbol{\theta}$ and $\boldsymbol{\theta}'$ within the rectangle $\mathcal{U}$, of size $L = \mathcal{O}(L_1, L_2)$, the next order term in (A17) is of order $L^9/|\mathbf{C}|^{10}$. We then note that only the coefficients $q_n$ depend on the center position $i, j$; hence, the sums of (A16) only need to be performed over the $q_n$, whereas the $v_n$ depend only on $\boldsymbol{\theta}$. It is clear that this summation can be carried out very quickly, even if the number of terms ($\sim M_x M_y$) is large. We have evaluated the kernel $\mathbf{H}(\boldsymbol{\theta}', \boldsymbol{\theta})$, using the preceding equations. Due to the explicit symmetry of our evaluation, the boundary condition (A2) is automatically satisfied at $\theta_1' = 0$ and at $\theta_2' = 0$; at the other two sides of the rectangle, the boundary condition is satisfied only if a sufficient number of terms are taken into account. The value of $\mathbf{H} \cdot \mathbf{n}$ at the two remaining sides thus provides a measure for the accuracy of the calculation. It turns out that using $N_x = N_y = 7$ in (A15) and $M_x = M_y = 400$ in (A16) yields an accuracy of better than $10^{-6}$.

The kernel $\mathbf{H}$ was calculated on a regular grid, with

$$\boldsymbol{\theta}_{ij}' = (i L_1/N_1, j L_2/N_2) \quad, \quad 0 \leq i \leq N_1 \quad, \quad 0 \leq j \leq N_2 \quad,$$

and

$$\boldsymbol{\theta}_{kl} = ((k-1/2)L_1/N_1, (l-1/2)L_2/N_2) \quad , \quad 1 \le k \le N_1 \quad , \quad 1 \le l \le N_2 \quad .$$

Of course, for a square it is useful to choose $N_1 = N_2$; otherwise, the gridcells should be chosen to be nearly quadratic. The numerical routine is very efficient and has on output the two arrays $H_1(\boldsymbol{\theta}'_{ij}, \boldsymbol{\theta}_{kl})$ and $H_2(\boldsymbol{\theta}'_{ij}, \boldsymbol{\theta}_{kl})$, i.e., arrays of dimension $(N_1 + 1) \times (N_2 + 1) \times N_1 \times N_2$. For example the computation time for $N_1 = N_2 = 50$ is about 20 minutes on an IBM risc 6000 processor.

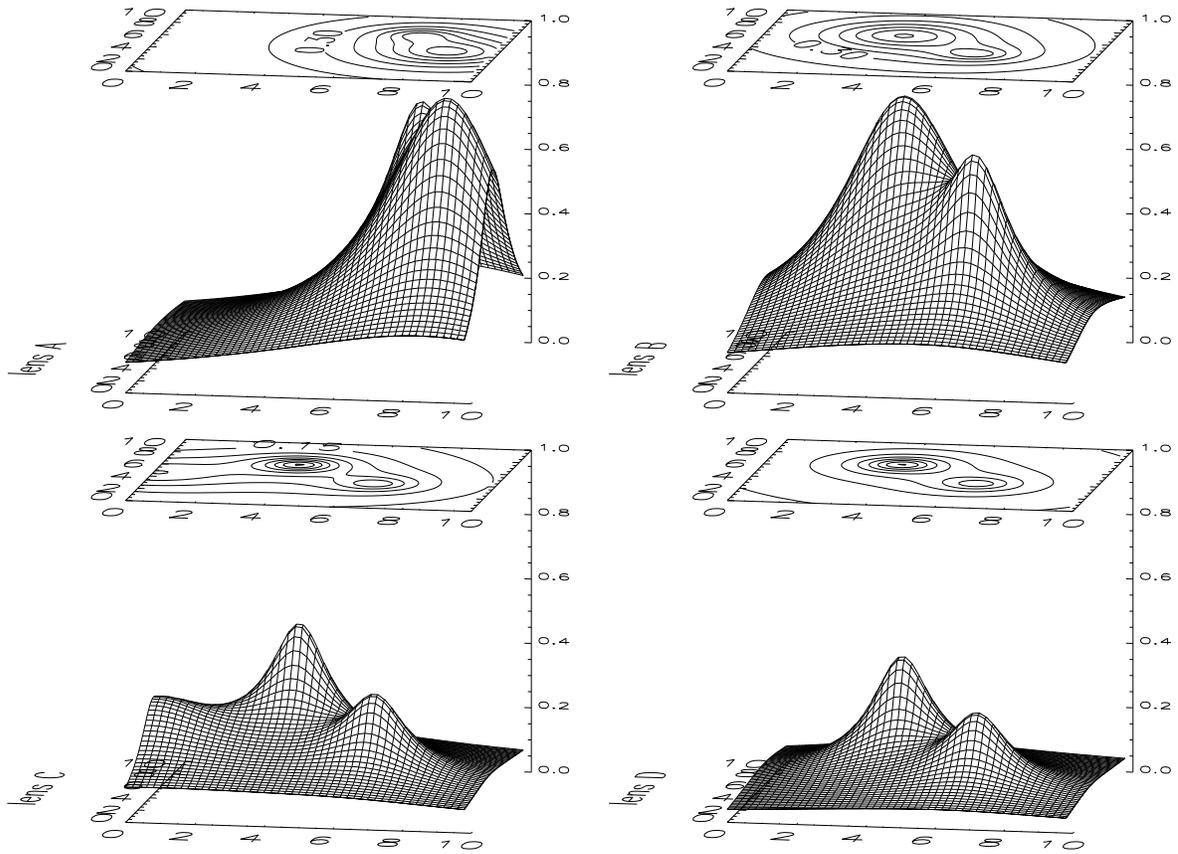

**Fig. 1.** Contour- and surface plot for the function $-K(\mathbf{x}) = -\ln\left[1 - \kappa(\mathbf{x})\right]$, where $\kappa$ is the two-dimensional surface mass density of the lens A,B,C and D respectively. In each panel, the size of the field is $7.'5 \times 7.'5$, and $K$ is calculated on a $50 \times 50$ grid; the spacing in the contour lines is 0.1 for lens A and lens B and 0.05 for lens C and D

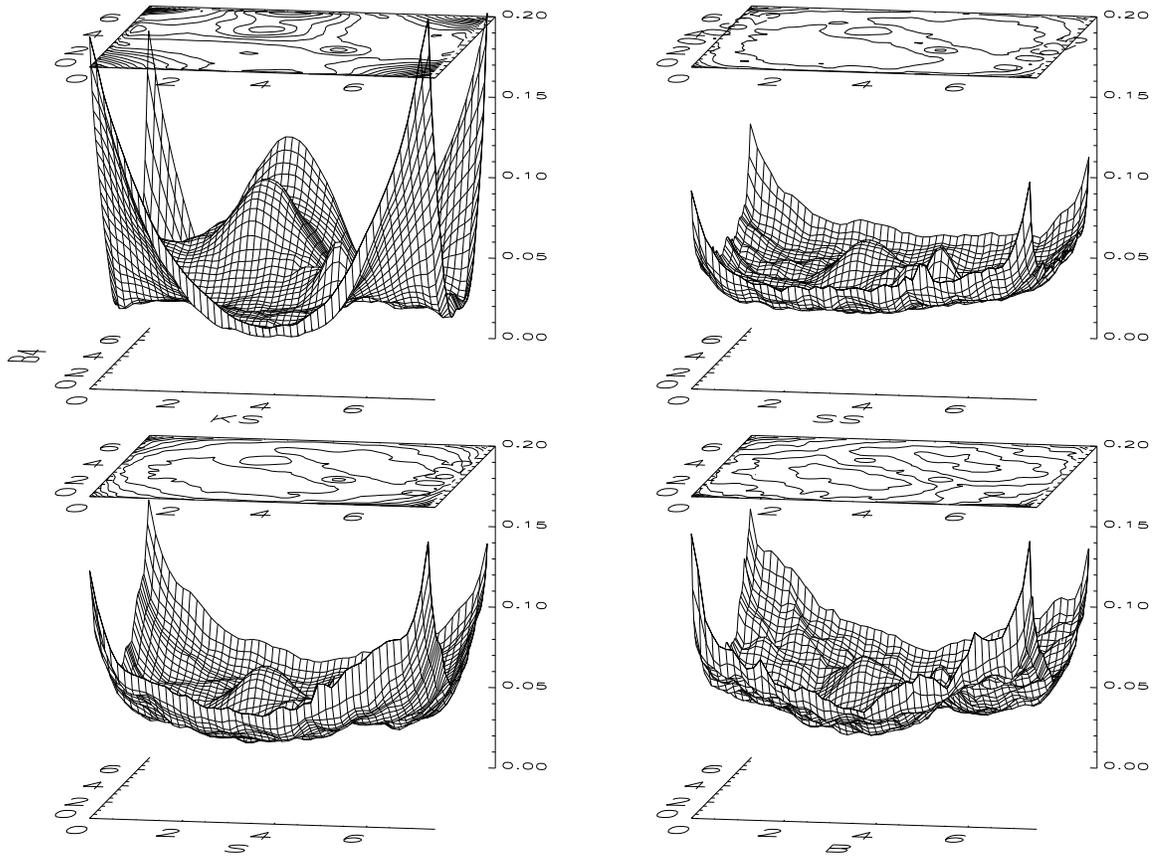

**Fig. 10.** Root mean square error of the reconstruction as a function of position for the KS-, SS-, S- and B-reconstruction. The number of simulations is 500, $n_0 = 50$, $\rho = 0.3$, $\theta_0 = 0.3$ and $\theta_{\rm sm} = 0.11$; the corresponding power spectra for the error field are displayed in Fig. 5b. The vertical axis is drawn from 0 to .2; the contour lines are in the same interval, with a step size of 0.0125; the size of the CCD field is 7.5 arcminutes, as usual

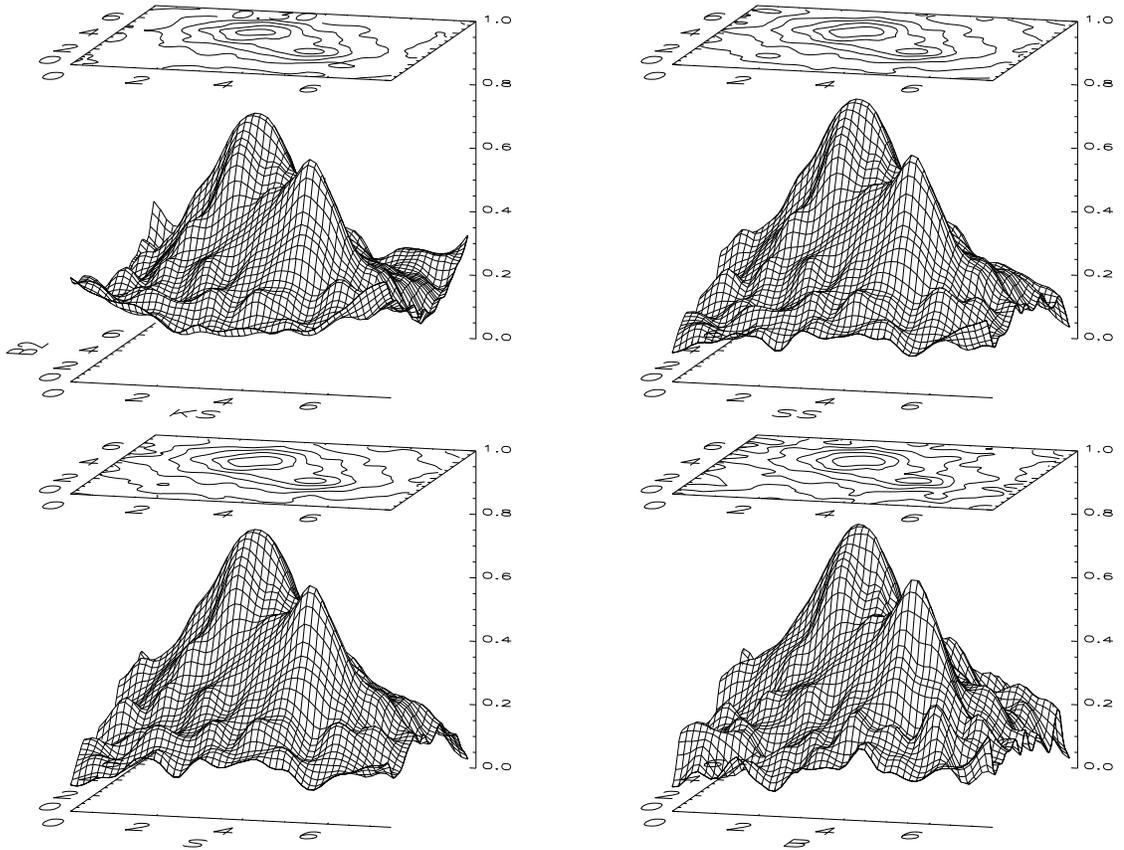

**Fig. 4.** Surface and contour plots of $-K^{\mathrm{KS}}$, $-K^{\mathrm{SS}}$, $-K^{\mathrm{S}}$ and $-K^{\mathrm{B}}$ obtained by the inversion of the lens B using the parameters of reconstruction B2, i.e. $L = 7.5$, $N = 50$, $n_0 = 50$, $\rho = 0.2$, $\theta_0 = 0.34$ and $\theta_{\mathrm{sm}} = 0.11$; the contours differ by 0.1